\documentclass[12pt,english,onecolumn]{IEEEtran}
\usepackage[T1]{fontenc}
\usepackage[latin1]{inputenc}
\usepackage{subfigure}
\usepackage{amsmath}
\usepackage{graphicx}
\usepackage{amssymb}

\makeatletter

\newcommand{\lyxdot}{.}

  \newtheorem{remrk}{Remark}
  \newtheorem{prop}{Proposition}
  \newtheorem{thm}{Theorem}
  \newtheorem{cor}{Corollary}
  \newtheorem{lemma}{Lemma}

\usepackage{cite}

\usepackage{bm}

\usepackage{bbm}

\usepackage{babel}
\makeatother
\begin{document}

\title{The Effect of Finite Rate Feedback on CDMA Signature Optimization
and MIMO Beamforming Vector Selection$^{*}$}

\author{Wei Dai, Youjian (Eugene) Liu, Brian Rider\\
 University of Colorado at Boulder\\
wei.dai@colorado.edu, eugeneliu@ieee.org, brian.rider@colorado.edu}

\maketitle
\begin{abstract}
We analyze the effect of finite rate feedback on CDMA (code-division
multiple access) signature optimization and MIMO (multi-input-multi-output)
beamforming vector selection. In CDMA signature optimization, for
a particular user, the receiver selects a signature vector from a
codebook to best avoid interference from other users, and then feeds
the corresponding index back to the specified user. For MIMO beamforming
vector selection, the receiver chooses a beamforming vector from a
given codebook to maximize throughput, and feeds back the corresponding
index to the transmitter. These two problems are dual: both can be
modeled as selecting a unit norm vector from a finite size codebook
to {}``match'' a randomly generated Gaussian matrix. In signature
optimization, the least match is required while the maximum match
is preferred for beamforming selection.

Assuming that the feedback link is rate limited, our main result is
an exact asymptotic performance formula where the length of the signature/beamforming
vector, the dimensions of interference/channel matrix, and the feedback
rate approach infinity with constant ratios. The proof rests on a
large deviation principle over a random matrix ensemble. Further,
we show that random codebooks generated from the isotropic distribution
are asymptotically optimal not only on average, but also with probability
one. 
\end{abstract}
\renewcommand{\thefootnote}{\fnsymbol{footnote}} \footnotetext[1]{The material in this paper was presented in part at the 40th Annual Conference on Information Sciences and Systems, March 2006. The work was supported in part by Thomson Inc. and by Junior Faculty Development Award, University of Colorado. Dai and Liu are with Department of Electrical and Computer Engineering, University of Colorado at Boulder. Rider is with Department of Mathematics, University of Colorado at Boulder. \\ This manuscipt was submitted to IEEE Trans. on Information Theory.} \renewcommand{\thefootnote}{\arabic{footnote}} \setcounter{footnote}{0}

\section{\label{sec:Introduction}Introduction}


In a direct-sequence code-division multiple access (DS-CDMA) system,
the performance is mainly limited by interference among users. We
assume that the receiver (base station) has perfect information of
all users' signature. For a particular user, the receiver selects
a signature to minimize the interference from other users, and then
feeds the corresponding index to the specified user through a feedback
link. Dually, consider a multi-input-multi-output (MIMO) system with
beamforming vector selection. Take the Rayleigh fading channel model,
where the channel matrix has independent and identically distributed
(i.i.d.) symmetric complex Gaussian entries with mean zero and unit
variance ($\mathcal{CN}\left(0,1\right)$). Now assume that the receiver
knows the channel state matrix perfectly. To aid the transmitter,
the receiver chooses a beamforming vector from the codebook to maximize
the throughput, and then feeds back the corresponding index to the
transmitter. In both scenarios, we consider a finite feedback rate
up to $R_{\mathrm{fb}}$ bits. Ideally, if the feedback rate is unlimited,
the transmitter is able to obtain interference/channel information
with arbitrary accuracy, but this is not practically feasible and
it is essential to real systems to understand the effect of finite
rate feedback.

This paper is the first to rigorously obtain 
exact asymptotic performance formulae for both problems when letting
the length of the signature/beamforming vector, the dimensions of
interference/channel matrix, and the feedback rate approach infinity
with constant ratios. The same set-ups had been considered previously
in \cite{Honig_Allerton2002_Limit_Feedback_Wireless_Channel} and
\cite{Honig_IT05_CDMA_signature_optimization_finite_feedback}, in
which a one-sided bound was presented (this was a lower bound on the
CDMA performance, and an upper bound in the case of MIMO). Our approach
is fundamentally different. Identifying the underlying problem as
a large deviation question for the connected random matrix ensemble,
we have a unified framework which handles both CDMA and MIMO cases
simultaneously%
\footnote{The analysis in \cite{Honig_IT05_CDMA_signature_optimization_finite_feedback}
is based on extreme order statistics, applied to the case of $n\uparrow\infty$
i.i.d. random variables with a \emph{fixed} distribution. The laws
of the underlying random variables for the problems at hand however
depend on $n$ in an essential way; attempting a proof through i.i.d.
order statistics results in needless complications.%
}. Further, while \cite{Honig_IT05_CDMA_signature_optimization_finite_feedback}
discusses the fact that random codebooks are asymptotically optimal
on average (their mean performance is the best achievable performance),
here we prove the stronger result that random codebooks are asymptotically
optimal with probability one.%
\footnote{We must add that, based on our earlier \cite{Dai_CISS06_CDMA_signature_optimization},
the authors of \cite{Honig_IT05_CDMA_signature_optimization_finite_feedback}
have gone on to refine their own estimates \cite{Honig_inprogress}.%
}

The paper is organized as follows. After describing the system models
in more detail, Section \ref{sec:Prelim} presents various needed
facts from Random Matrix Theory. Section \ref{sec:Main-Results} contains
the main results. The basic convergence result is Theorem \ref{thm:main-theorem-performance},
which in turn is based on a random codebook version, Theorem \ref{thm:Average-Performance-Random-Codes},
along with a separate argument that any given codebook will not asymptotically
outperform its random counterpart. This section concludes with the
almost sure optimality, Theorem \ref{thm:Optimality-Random-Codes}.
Once again, all this is based on a large deviation principle for the
spectrum of a Wishart type random matrix. That proof is found in the
appendices.

\begin{remrk}
Our methods carry over to the problem of the average throughput of
an MMSE receiver in CDMA systems. Each appearance of $\frac{1}{n}\mathbf{H}\mathbf{H}^{\dagger}$,
in say (\ref{eq:Def-c-max-n}) below, is replaced by $\left(\mathbf{I}+\frac{1}{n}\mathbf{H}\mathbf{H}^{\dagger}\right)^{-1}$,
and the proof may be followed verbatim except for the few obvious
(and trivial) modifications. 
\end{remrk}

\section{\label{sec:System-Model}System Model}


\subsection{\label{sub:CDMA-Signature-Optimization}CDMA Signature Optimization}

$ $

In a sampled discrete-time symbol-synchronous DS-CDMA system with
$m$ users, the received vector $\mathbf{Y}$ can be written as

\[
\mathbf{Y}=\sum_{j=1}^{m}B_{j}\mathbf{s}_{j}+\mathbf{W},\]
 where $B_{j}\in\mathbb{C}$ and $\mathbf{s}_{j}\in\mathbb{C}^{n\times1}$
are the transmitted symbol and the signature vector for user $j$
respectively, and $\mathbf{W}\in\mathbb{C}^{n\times1}$ is the additive
white Gaussian noise vector with zero mean and covariance matrix $\sigma^{2}\mathbf{I}$.
Throughout this paper, we assume that the transmitted symbols $B_{j}$'s
are i.i.d. $\mathcal{CN}\left(0,1\right)$ random variables. The signature
vectors $\mathbf{s}_{j}$'s satisfy $\mathbf{s}_{j}^{\dagger}\mathbf{s}_{j}=1$,
$1\le j\le m$. Their length $n$ is often referred to as processing
gain in literature.

This paper focuses on matched filter receiver. As already mentioned,
the analysis for a MMSE receiver is effectively the same. With a matched
filter receiver, the throughput of user $1$ is \[
\log\left(1+\frac{1}{\sigma^{2}+\sum_{j=2}^{m}\mathbf{s}_{1}^{\dagger}\mathbf{S}_{1}\mathbf{S}_{1}^{\dagger}\mathbf{s}_{1}}\right),\]
 where $\mathbf{S}_{1}=\left[\mathbf{s}_{2}\cdots\mathbf{s}_{m}\right]$.

The signature optimization is described as follows. Assume that the
receiver has perfect knowledge of the $\mathbf{s}_{j}$'s. It guides
a particular user, say user 1, to avoid the others' interference.
Here, a codebook $\mathcal{B}$ of signature vectors is declared to
both the receiver and user 1. Given the other users' signatures $\mathbf{s}_{2},\cdots,\mathbf{s}_{m}$,
the receiver selects\[
\mathbf{s}_{1}=\underset{\mathbf{v}\in\mathcal{B}}{\arg\;\min}\;\mathbf{v}^{\dagger}\mathbf{S}_{1}\mathbf{S}_{1}^{\dagger}\mathbf{v}.\]
 Then it feeds the corresponding index back to user $1$ through a
finite rate feedback link, whose rate is up to $R_{\mathrm{fb}}$
bits. This finite feedback rate assumption imposes a constraint on
the size of the codebook, $\left|\mathcal{B}\right|\le2^{R_{\mathrm{fb}}}$.
Therefore, the average interference for user $1$ is given by \[
\underset{\mathcal{B}:\;\left|\mathcal{B}\right|\le2^{R_{\mathrm{fb}}}}{\inf}\mathrm{E}_{\mathbf{S}_{1}}\left[\underset{\mathbf{v}\in\mathcal{B}}{\min}\;\mathbf{v}^{\dagger}\mathbf{S}_{1}\mathbf{S}_{1}^{\dagger}\mathbf{v}\right].\]

\subsection{\label{sub:MIMO-Beamforming-Vector}MIMO Beamforming Vector Selection}

$ $

The signal model for a MIMO system with beamforming vector selection
is \[
\mathbf{Y}=\mathbf{H}^{\dagger}\mathbf{q}X+\mathbf{W},\]
 where $\mathbf{Y}\in\mathbb{C}^{m\times1}$ is the received signal
vector, $\mathbf{H}\in\mathbb{C}^{n\times m}$ is the channel state
matrix, $\mathbf{q}\in\mathbb{C}^{n\times1}$ is the beamforming vector
satisfying $\mathbf{q}^{\dagger}\mathbf{q}=1$, $X$ is the transmitted
signal $\mathcal{CN}\left(0,1\right)$, $\mathbf{W}\in\mathbb{C}^{m\times1}$
is the white Gaussian noise vector with mean zero and covariance $\sigma^{2}\mathbf{I}_{m}$.
The dimensions $n$ and $m$ are the numbers of antennas at the transmitter
and receiver.

In the above setting, beamforming vector selection proceeds as follows.
Assume that the receiver knows the realization of $\mathbf{H}$ perfectly,
and feeds beamforming vector selection information back to the transmitter
through a feedback link with rate up to $R_{\mathrm{fb}}$bits. A
codebook $\mathcal{B}$ containing $2^{R_{\mathrm{fb}}}$ many candidate
beamforming vector is declared to both transmitter and receiver. For
any $\mathbf{H}$ realization, the receiver selects the beamforming
vector to maximize the throughput \begin{align*}
\mathbf{q} & =\underset{\mathbf{v}\in\mathcal{B}}{\arg\;\max}\;\log\left|\mathbf{I}_{m}+\frac{1}{\sigma^{2}}\mathbf{H}^{\dagger}\mathbf{v}\mathbf{v}^{\dagger}\mathbf{H}\right|\\
 & =\underset{\mathbf{v}\in\mathcal{B}}{\arg\;\max}\;\mathbf{v}^{\dagger}\left(\frac{1}{n}\mathbf{H}\mathbf{H}^{\dagger}\right)\mathbf{v}.\end{align*}
 The corresponding index is fed back to the transmitter, which then
employs $\mathbf{q}$ for transmission. The average received signal
power is \[
\underset{\mathcal{B}:\;\left|\mathcal{B}\right|\le2^{R_{\mathrm{fb}}}}{\sup}\mathrm{E}_{\mathbf{H}}\left[\underset{\mathbf{v}\in\mathcal{B}}{\max}\;\mathbf{v}^{\dagger}\left(\frac{1}{n}\mathbf{H}\mathbf{H}^{\dagger}\right)\mathbf{v}\right].\]

\subsection{\label{sub:Preliminaries}Unified Formulation}

$ $

It is difficult to quantify both the average interference in Section
\ref{sub:CDMA-Signature-Optimization} and the average received power
in Section \ref{sub:MIMO-Beamforming-Vector}. However, when $n$,
$m$ and $R_{\mathrm{f}b}$ approach infinity linearly with constant
ratios, each converges to a constant. To be precise, let $\mathcal{B}\triangleq\left\{ \mathbf{v}\in\mathbb{C}^{n\times1}:\;\mathbf{v}^{\dagger}\mathbf{v}=1\right\} $
be a codebook. Let $\mathbf{H}\in\mathbb{C}^{n\times m}$ be a random
Gaussian matrix with i.i.d. $\mathcal{CN}\left(0,1\right)$ entries.
Define \begin{equation}
c_{\min,n}=\underset{\mathcal{B}:\;\left|\mathcal{B}\right|=2^{R_{\mathrm{fb}}}}{\inf}\mathrm{E}_{\mathbf{H}}\left[\underset{\mathbf{v}\in\mathcal{B}}{\min}\;\mathbf{v}^{\dagger}\left(\frac{1}{n}\mathbf{H}\mathbf{H}^{\dagger}\right)\mathbf{v}\right]\label{eq:Def-c-min-n}\end{equation}
 and \begin{equation}
c_{\max,n}=\underset{\mathcal{B}:\;\left|\mathcal{B}\right|=2^{R_{\mathrm{fb}}}}{\sup}\mathrm{E}_{\mathbf{H}}\left[\underset{\mathbf{v}\in\mathcal{B}}{\max}\;\mathbf{v}^{\dagger}\left(\frac{1}{n}\mathbf{H}\mathbf{H}^{\dagger}\right)\mathbf{v}\right].\label{eq:Def-c-max-n}\end{equation}
 As $n,m,R_{\mathrm{fb}}\rightarrow\infty$ with with $\frac{m}{n}\rightarrow\frac{1}{\beta}\in\mathbb{R}^{+}$
and $\frac{R_{\mathrm{fb}}}{n}\rightarrow r\in\mathbb{R}^{+}$, we
shall show that $c_{\min,n}$ and $c_{\max,n}$ converge to constants
and compute their limits in Section \ref{sec:Main-Results}.

\begin{remrk}
We assume that $\mathbf{H}\in\mathbb{C}^{n\times m}$ has i.i.d. entries
in this unified formulation while the matrix $\mathbf{S}_{1}\in\mathbb{C}^{n\times\left(m-1\right)}$
in the CDMA signature optimization is composed of independent and
isotropically distributed columns. Notably, the asymptotic statistics
of $\mathbf{S}_{1}\mathbf{S}_{1}^{\dagger}$ and $\frac{1}{n}\mathbf{H}\mathbf{H}^{\dagger}$
are the same as $\frac{m}{n}\rightarrow\frac{1}{\beta}\in\mathbb{R}$.
The limit of $c_{\min,n}$ will gives the asymptotic average interference
for user 1 in a CDMA system. 
\end{remrk}

\section{\label{sec:Prelim}Preliminaries}

\subsection{Asymptotic Random Matrix Theory}

$ $

The performance calculation is based on the asymptotic spectral distribution
of the matrix $\frac{1}{m}\mathbf{H}\mathbf{H}^{\dagger}$. Let $\lambda_{1},\cdots,\lambda_{n}$
be the $n$ singular values of $\frac{1}{m}\mathbf{H}\mathbf{H}^{\dagger}$.
Define the empirical distribution of the singular values \[
\mu_{n,\bm{\lambda}}\left(\lambda\right)\triangleq\frac{1}{n}\left|\left\{ j:\;\lambda_{j}\le\lambda\right\} \right|.\]
 As $n,m\rightarrow\infty$ with $\frac{m}{n}\rightarrow\frac{1}{\beta}\in\mathbb{R}^{+}$,
\begin{equation}
d\mu_{\lambda}=\underset{\left(n,m\right)\rightarrow\infty}{\lim}d\mu_{n,\bm{\lambda}}\left(\lambda\right)=\left(\left(1-\frac{1}{\beta}\right)^{+}\delta\left(\lambda\right)+\frac{\sqrt{\left(\lambda-\lambda^{-}\right)^{+}\left(\lambda^{+}-\lambda\right)^{+}}}{2\pi\beta\lambda}\right)\, d\lambda\label{eq:spectrum-pdf}\end{equation}
 almost surely, where $\lambda^{\pm}=\left(1\pm\sqrt{\beta}\right)^{2}$
and $\left(x\right)^{+}=\max\left(x,0\right)$. (A good reference
for this type of result is \cite{Silverstein_1995_strong_convergence}.)
For later it will be useful to define\[
\lambda_{t}^{-}\triangleq\begin{cases}
0 & \mathrm{if}\;\beta\ge1\\
\lambda^{-} & \mathrm{if}\;\beta<1\end{cases},\;\mathrm{and}\;\bar{\lambda}=\int\lambda\cdot d\mu_{\lambda}.\]
 Consider as well a linear spectral statistic \[
g\left(\frac{1}{m}\mathbf{H}\mathbf{H}^{\dagger}\right)=\frac{1}{n}\sum_{i=1}^{n}g\left(\lambda_{i}\right).\]
 If $g$ is Lipschitz on $\left[\lambda_{t}^{-},\lambda^{+}\right]$,
then we also have that \[
\underset{\left(n,m\right)\rightarrow\infty}{\lim}g\left(\frac{1}{m}\mathbf{H}\mathbf{H}^{\dagger}\right)=\int g\left(\lambda\right)d\mu_{\lambda}\]
 almost surely, see for example \cite{Guionnet_Zeitouni_2000} for
a modern approach.

Last, the asymptotic properties of the minimum and maximum eigenvalues
will figure into our analysis. For any finite $n$, set $\lambda_{\min,\bm{\lambda}}=\underset{1\le i\le n}{\min}\lambda_{i}$
and $\lambda_{\max,\bm{\lambda}}=\underset{1\le i\le n}{\max}\lambda_{i}$.

\begin{prop}
\label{prop:Asymptotic-Lambda-Max}Let $n,m\rightarrow\infty$ linearly
with $\frac{m}{n}\rightarrow\frac{1}{\beta}\in\mathbb{R}^{+}$. 
\begin{enumerate}
\item $\lambda_{\min,n}\rightarrow\lambda_{t}^{-}$ and $\lambda_{\max,n}\rightarrow\lambda^{+}$
almost surely. 
\item All moments of $\lambda_{\min,n}$ and $\lambda_{\max,n}$ also converge. 
\end{enumerate}
\end{prop}
The almost sure convergence goes back to \cite{Bai_Yin_Krishnaih_1988,Bai_Yin_1993}.
The convergence of moments is implied by the results in \cite{Ledoux_2005_Orthogonal_Polynomials}.
A direct application of this proposition is that for $\forall A_{n}\subset\mathbb{R}^{n}$
such that $\mu_{n,\bm{\lambda}}\left(A_{n}\right)\rightarrow0$, $\mathrm{E}_{\bm{\lambda}}\left[\lambda_{\max,\bm{\lambda}},\; A_{n}\right]\rightarrow0$;
this fact will be employed repeatedly below.

\subsection{Isotropic Distribution}

$ $

We also bring in the isotropic distribution for \[
\mathcal{U}_{n\times m}\triangleq\left\{ \mathbf{U}\in\mathbb{C}^{n\times m}:\;\mathbf{U}^{\dagger}\mathbf{U}=\mathbf{I}_{m}\right\} ,\]
 by which we mean the (left) Haar measure $\mu$ of $\mathcal{U}_{n\times m}$.
In particular, for any set $\mathcal{M}\subset\mathcal{U}_{n\times m}$
and any $\mathbf{U}\in\mathcal{U}_{n\times n}$, $\mu\left(\mathbf{U}\mathcal{M}\right)=\mu\left(\mathcal{M}\right)$.

\section{\label{sec:Main-Results}Main Results}

For $\forall x\in\left(\lambda_{t}^{-},\lambda^{+}\right)$, define\begin{align}
\psi_{x}\left(\alpha\right) & \triangleq\begin{cases}
-\int\log\left(1-\alpha\left(\lambda-x\right)\right)d\mu_{\lambda} & \mathrm{if}\;\alpha\in\left[-\frac{1}{x-\lambda_{t}^{-}},\frac{1}{\lambda^{+}-x}\right]\\
+\infty & \mathrm{otherwise}\end{cases}\label{eq:moment-generate-fn-def}\end{align}
 and\begin{equation}
\psi_{x}^{*}\left(t\right)\triangleq\underset{\alpha\in\mathbb{R}}{\sup}\;\alpha t-\psi_{x}\left(\alpha\right)\label{eq:rate-function-def}\end{equation}
 for any $t\in\mathbb{R}$. Our basic convergence result for $c_{\min,n}$
and $c_{\max,n}$ reads:

\begin{thm}
\label{thm:main-theorem-performance}Let $n$, $m$ and $R_{\mathrm{fb}}$
approach infinity linearly with $\frac{m}{n}\rightarrow\frac{1}{\beta}\in\mathbb{R}^{+}$
and $\frac{R_{\mathrm{fb}}}{n}\rightarrow r\in\mathbb{R}^{+}$. There
exist unique $x_{r}^{-}\in\left(\lambda_{t}^{-},\bar{\lambda}\right)$
and $x_{r}^{+}\in\left(\bar{\lambda},\lambda^{+}\right)$ such that
$r\log2=\psi_{x_{r}^{-}}^{*}\left(0\right)=\psi_{x_{r}^{+}}^{*}\left(0\right)$.
Furthermore, \[
\underset{\left(n,m,R_{\mathrm{fb}}\right)\rightarrow\infty}{\lim}\; c_{\min,n}=x_{r}^{-}/\beta,\]
 and\[
\underset{\left(n,m,R_{\mathrm{fb}}\right)\rightarrow\infty}{\lim}\; c_{\max,n}=x_{r}^{+}/\beta.\]

\end{thm}

\begin{remrk}
From the properties of $\psi_{x}^{*}\left(0\right)$ (Proposition
\ref{pro:Psi^Star-x}), \[
\begin{cases}
c_{\min,n}\rightarrow\bar{\lambda}/\beta,\; c_{\max,n}\rightarrow\bar{\lambda}/\beta & \mathrm{as}\; r\downarrow0\\
c_{\min,n}\rightarrow\lambda_{t}^{-}/\beta,\; c_{\max,n}\rightarrow\lambda^{+}/\beta\; & \mathrm{as}\; r\uparrow\infty\end{cases}.\]
 This is consistent with intuition: $r=0$ and $r=\infty$ representing
either no, or perfect information. 

\end{remrk}
Using ideas from \cite{Verdu_IT1999_CDMA_Spectral_Efficiency}, we
may also obtain fairly explicit formulas for $x_{r}^{-}$ and $x_{r}^{+}$.
(It should be noted that \cite{Honig_inprogress} also takes on this
computation, but from a different vantage point.)

\begin{cor}
\label{cor:main-result-computations}Let $r_{\min}=\frac{-\log\left(1-\sqrt{\beta}\right)-\sqrt{\beta}}{\beta\log2}$
for $\forall\beta<1$ and $r_{\max}=\frac{\sqrt{\beta}-\log\left(1+\sqrt{\beta}\right)}{\beta\log2}$
for $\forall\beta\in\mathbb{R}^{+}$. Then for any $r\in\mathbb{R}^{+}$,
$x_{r}^{-}\in\left(\lambda_{t}^{-},1\right)$ satisfies \[
\begin{cases}
x_{r}^{-}=\left(1-\sqrt{\beta}\right)^{2}+\sqrt{\beta}\left(1-\sqrt{\beta}\right)^{1-\frac{1}{\beta}}\exp\left(-\frac{1}{\sqrt{\beta}}-r\log2\right) & \mathrm{if}\;\beta<1\;\mathrm{and}\; r>r_{\min}\\
x_{r}^{-}=e^{x_{r}^{-}-1}2^{-\beta r} & \mathrm{otherwise}\end{cases},\]
 and $x_{r}^{+}\in\left(1,\lambda^{+}\right)$ satisfies \[
\begin{cases}
x_{r}^{+}=\left(1+\sqrt{\beta}\right)^{2}-\sqrt{\beta}\left(1+\sqrt{\beta}\right)^{1-\frac{1}{\beta}}\exp\left(\frac{1}{\sqrt{\beta}}-r\log2\right) & \mathrm{if}\; r>r_{\max}\\
x_{r}^{+}=e^{x_{r}^{+}-1}2^{-\beta r} & \mathrm{otherwise}\end{cases}.\]

\end{cor}

Granted the existence and uniqueness of $x_{r}^{-}$ and $x_{r}^{+}$,
which follow from basic properties of $\psi_{x}^{*}\left(0\right)$
established in Proposition \ref{pro:Psi^Star-x} of Appendix \ref{sec:Properties-of-Rate-Functions},
the proof of Theorem \ref{thm:main-theorem-performance} takes the
following course. First, by calculating the average performance of
random codes, we are construct upper and lower bounds on $\lim c_{\min,n}$
and $\lim c_{\max,n}$ respectively. Let $\mathcal{B}_{\mathrm{rand}}$
be a randomly constructed codebook of i.i.d. unit-norm vectors from
the isotropic distribution. In particular, $\mathcal{B}_{\mathrm{rand}}=\left\{ \mathbf{v}_{1},\cdots,\mathbf{v}_{2^{R_{\mathrm{fb}}}}\right\} $,
where $\mathbf{v}_{k}=\mathbf{z}_{k}/\left\Vert \mathbf{z}_{k}\right\Vert $,
$\mathbf{z}_{k}=\left[z_{k,1},\cdots,z_{k,n}\right]^{\dagger}$ and
$z_{k,i}$ are i.i.d. $\mathcal{CN}\left(0,1\right)$ for all $1\leq k\leq2^{R_{\mathrm{fb}}}$
and $1\leq i\leq n$. Define\[
c_{\min,n,\mathrm{rand}}=\mathrm{E}_{\mathcal{B}_{\mathrm{rand}}}\left[\mathrm{E}_{\mathbf{H}}\left[\underset{\mathbf{v}\in\mathcal{B}_{\mathrm{rand}}}{\min}\;\mathbf{v}^{\dagger}\left(\frac{1}{n}\mathbf{H}\mathbf{H}^{\dagger}\right)\mathbf{v}\right]\right]\]
 and\[
c_{\max,n,\mathrm{rand}}=\mathrm{E}_{\mathcal{B}_{\mathrm{rand}}}\left[\mathrm{E}_{\mathbf{H}}\left[\underset{\mathbf{v}\in\mathcal{B}_{\mathrm{rand}}}{\max}\;\mathbf{v}^{\dagger}\left(\frac{1}{n}\mathbf{H}\mathbf{H}^{\dagger}\right)\mathbf{v}\right]\right].\]
 The following theorem calculates the average performance of random
codes.

\begin{thm}
\label{thm:Average-Performance-Random-Codes}As $n,m,R_{\mathrm{fb}}\rightarrow\infty$
with $\frac{m}{n}\rightarrow\frac{1}{\beta}\in\mathbb{R}^{+}$ and
$\frac{R_{\mathrm{fb}}}{n}\rightarrow r\in\mathbb{R}^{+}$, \[
\underset{\left(n,m,R_{\mathrm{fb}}\right)\rightarrow\infty}{\lim}\; c_{\min,n,\mathrm{rand}}=x_{r}^{-}/\beta\]
 and\[
\underset{\left(n,m,R_{\mathrm{fb}}\right)\rightarrow\infty}{\lim}\; c_{\max,n,\mathrm{rand}}=x_{r}^{+}/\beta.\]

\end{thm}

Clearly, $\lim\; c_{\min,n}\le\lim\; c_{\min,n,\mathrm{rand}}$ and
$\lim\; c_{\max,n}\ge\lim\; c_{\max,n,\mathrm{rand}}$, and the next
step is to obtain a lower bound $\lim\; c_{\min,n}$ and and upper
bound $\lim\; c_{\max,n}$. Introduce the singular value decomposition,
$\frac{1}{m}\mathbf{H}\mathbf{H}^{\dagger}=\mathbf{U}\mathbf{\Lambda}\mathbf{U}^{\dagger}$
where $\mathbf{U}\in\mathcal{U}_{n\times n}$ and $\mathbf{\Lambda}\in\mathbb{R}^{n\times n}$
is the diagonal matrix of eigenvalues $\lambda_{1},\cdots,\lambda_{n}$.
It is well known that $\mathbf{U}$ is isotropically distributed and
independent with $\mathbf{\Lambda}$. For any codebook $\mathcal{B}=\left\{ \mathbf{v}\in\mathcal{U}_{n\times1}\right\} $,
define \begin{equation}
c_{\min,n,\bm{\lambda},\mathcal{B}}\triangleq\mathrm{E}_{\mathbf{U}}\left[\left.\underset{\mathbf{v}\in\mathcal{B}}{\min}\;\mathbf{v}^{\dagger}\mathbf{U}\mathrm{diag}\left(\bm{\lambda}\right)\mathbf{U}^{\dagger}\mathbf{v}\right|\bm{\lambda}\right]\label{eq:c_min_n_arbitrary_code}\end{equation}
 and \begin{equation}
c_{\max,n,\bm{\lambda},\mathcal{B}}\triangleq\mathrm{E}_{\mathbf{U}}\left[\left.\underset{\mathbf{v}\in\mathcal{B}}{\max}\;\mathbf{v}^{\dagger}\mathbf{U}\mathrm{diag}\left(\bm{\lambda}\right)\mathbf{U}^{\dagger}\mathbf{v}\right|\bm{\lambda}\right].\label{eq:c_max_n_arbitrary_code}\end{equation}
 As $n,m,R_{\mathrm{fb}}\rightarrow\infty$ linearly with constant
ratios $\frac{1}{\beta}$ and $r$, Define \[
c_{\min,\bm{\lambda}}\triangleq\underset{\left(n,m,R_{\mathrm{fb}}\right)\rightarrow\infty}{\lim}\;\underset{\mathcal{B}:\;\left|\mathcal{B}\right|=2^{R_{\mathrm{fb}}}}{\inf}\; c_{\min,n,\bm{\lambda},\mathcal{B}}\]
 and \[
c_{\max,\bm{\lambda}}\triangleq\underset{\left(n,m,R_{\mathrm{fb}}\right)\rightarrow\infty}{\lim}\;\underset{\mathcal{B}:\;\left|\mathcal{B}\right|=2^{R_{\mathrm{fb}}}}{\inf}\; c_{\max,n,\bm{\lambda},\mathcal{B}}.\]
 It is clear that $c_{\min,\bm{\lambda}}$ and $c_{\max,\bm{\lambda}}$
are random variables depending on $\bm{\lambda}$. The following theorem
provides bounds on $c_{\min,\bm{\lambda}}$ and $c_{\max,\bm{\lambda}}$,
and therefore bounds on $\lim\; c_{\min,n}$ and $\lim\; c_{\max,n}$.

\begin{thm}
\label{thm:asymptotic-bds-all-codebooks}As $n,\; m,\; R_{\mathrm{fb}}\rightarrow\infty$
with $\frac{m}{n}\rightarrow\frac{1}{\beta}\in\mathbb{R}^{+}$ and
$\frac{R_{\mathrm{fb}}}{n}\rightarrow r\in\mathbb{R}^{+}$, 
\begin{enumerate}
\item $c_{\min,\bm{\lambda}}\ge x_{r}^{-}$ and $c_{\max,\bm{\lambda}}\le x_{r}^{+}$
with probability 1 in $\bm{\lambda}$, and 
\item $\underset{\left(n,m,R_{\mathrm{fb}}\right)\rightarrow\infty}{\lim}c_{\min,n}\ge x_{r}^{-}/\beta$
and $\underset{\left(n,m,R_{\mathrm{fb}}\right)\rightarrow\infty}{\lim}c_{\max,n}\le x_{r}^{+}/\beta$. 
\end{enumerate}
\end{thm}
By combining the above results, Theorem \ref{thm:main-theorem-performance}
is proved.

Finally, while Theorem \ref{thm:Average-Performance-Random-Codes}
implies that random codebooks are asymptotically optimal on average,
we actually have the stronger result that they are asymptotically
optimal with probability one.

\begin{thm}
\label{thm:Optimality-Random-Codes}As $n,\; m,\; R_{\mathrm{fb}}\rightarrow\infty$
with $\frac{m}{n}\rightarrow\frac{1}{\beta}\in\mathbb{R}^{+}$ and
$\frac{R_{\mathrm{fb}}}{n}\rightarrow r\in\mathbb{R}^{+}$, for any
$\epsilon>0$\[
\underset{\left(n,m,R_{\mathrm{fb}}\right)\rightarrow\infty}{\lim}\mu_{n,\mathcal{B}_{\mathrm{rand}}}\left(\mathrm{E}_{\mathbf{H}}\left[\underset{\mathbf{v}\in\mathcal{B}_{\mathrm{rand}}}{\min}\frac{1}{n}\mathbf{v}^{\dagger}\mathbf{H}\mathbf{H}^{\dagger}\mathbf{v}\right]>\beta\cdot x_{r}^{-}+\epsilon\right)=0\]
 and\[
\underset{\left(n,m,R_{\mathrm{fb}}\right)\rightarrow\infty}{\lim}\mu_{n,\mathcal{B}_{\mathrm{rand}}}\left(\mathrm{E}_{\mathbf{H}}\left[\underset{\mathbf{v}\in\mathcal{B}_{\mathrm{rand}}}{\max}\frac{1}{n}\mathbf{v}^{\dagger}\mathbf{H}\mathbf{H}^{\dagger}\mathbf{v}\right]<\beta\cdot x_{r}^{+}-\epsilon\right)=0.\]

\end{thm}

\begin{remrk}
\label{rem:Throughput}The asymptotic achievable throughputs of the
above CDMA and MIMO systems are \[
\log\left(1+\frac{1}{\sigma^{2}+\lim\; c_{\min,n}}\right)\]
 and \[
\log\left(1+\frac{\lim\; c_{\max,n}}{\sigma^{2}}\right)\]
 respectively. These facts are direct applications of the proof of
Theorem \ref{thm:asymptotic-bds-all-codebooks}.
\end{remrk}

The proofs of Theorem \ref{thm:Average-Performance-Random-Codes}-\ref{thm:Optimality-Random-Codes}
occupy the next sections (\ref{sub:average-performance-random-codes}-\ref{sub:Optimality-Random-Codes}).
The key step is a large deviation principle established in Theorem
\ref{thm:LD-Y} in Appendix \ref{sec:Large-Deviation-Principles}.
Last, the computation in Corollary \ref{cor:main-result-computations}
is conducted in Appendix \ref{sec:Computation}.

\subsection{\label{sub:average-performance-random-codes}Average Performance
of Random Codes}

$ $

Since the calculations of $c_{\max,n,\mathrm{rand}}$ and $c_{\min,n,\mathrm{rand}}$
follow the same line, we only give the details for $c_{\min,n,\mathrm{rand}}$.
In the following, we first prove that \begin{equation}
\underset{\left(n,m,R_{\mathrm{fb}}\right)\rightarrow\infty}{\lim}\; c_{\min,n,\mathrm{rand}}\ge x_{r}^{-}/\beta\label{eq:c_rand_lb}\end{equation}
 by Chebyshev's inequality, then show that \begin{equation}
\underset{\left(n,m,R_{\mathrm{fb}}\right)\rightarrow\infty}{\lim}\; c_{\min,n,\mathrm{rand}}\le x_{r}^{-}/\beta\label{eq:c_rand_ub}\end{equation}
 by exponential change of a probability measure.

We express $c_{\min,n,\mathrm{rand}}$ in a convenient form. Recall
the singular value decomposition $\frac{1}{m}\mathbf{H}\mathbf{H}^{\dagger}=\mathbf{U}\mathbf{\Lambda}\mathbf{U}^{\dagger}$.
\begin{align*}
c_{\min,n,\mathrm{rand}} & =\mathrm{E}_{\mathbf{H}}\left[\mathrm{E}_{\mathcal{B}_{\mathrm{rand}}}\left[\frac{1}{n}\underset{k}{\min}\frac{\mathbf{z}_{k}^{\dagger}\mathbf{H}\mathbf{H}^{\dagger}\mathbf{z}_{k}}{\left\Vert \mathbf{z}_{k}\right\Vert ^{2}}\right]\right]\\
 & =\frac{m}{n}\mathrm{E}_{\mathbf{H}}\left[\mathrm{E}_{\mathcal{B}_{\mathrm{rand}}}\left[\underset{k}{\min}\frac{\mathbf{z}_{k}^{\dagger}\left(\frac{1}{m}\mathbf{H}\mathbf{H}^{\dagger}\right)\mathbf{z}_{k}}{\left\Vert \mathbf{z}_{k}\right\Vert ^{2}}\right]\right]\\
 & =\frac{m}{n}\mathrm{E}_{\mathbf{H}}\left[\mathrm{E}_{\mathcal{B}_{\mathrm{rand}}}\left[\underset{k}{\min}\frac{\mathbf{z}_{k}^{\dagger}\mathbf{U}\mathbf{\Lambda}\mathbf{U}^{\dagger}\mathbf{z}_{k}}{\left\Vert \mathbf{z}_{k}\right\Vert ^{2}}\right]\right]\\
 & =\frac{m}{n}\mathrm{E}_{\mathbf{H}}\left[\mathrm{E}_{\mathcal{B}_{\mathrm{rand}}}\left[\underset{k}{\min}\frac{\sum_{i=1}^{n}\lambda_{i}\left|z_{k,i}\right|^{2}}{\sum_{i=1}^{n}\left|z_{k,i}\right|^{2}}\right]\right],\end{align*}
 where the last equality follows from the fact that $\mathbf{z}_{k}$
and $\mathbf{U}\mathbf{z}_{k}$ are statistically equal for any given
$n\times n$ unitary matrix $\mathbf{U}$. Let $Y_{k,i}=\left|Z_{k,i}\right|^{2}$.
Then $Y_{k,i}$'s ($1\le k\le2^{R_{\mathrm{fb}}}$ and $1\le i\le n$)
are i.i.d. random variables with probability measure $d\mu_{y}=e^{-y}dy$.
Note that for a given $\bm{\lambda}$ vector, the random variables
$\sum_{i=1}^{n}\lambda_{i}Y_{k,i}/\sum_{i=1}^{n}Y_{k,i}$'s ($k=1,\cdots,2^{R_{\mathrm{fb}}}$)
are conditional independent (conditioned on $\bm{\lambda}$). Define
the corresponding conditional probability measure \begin{align}
\mu_{n,\mathbf{Y}}\left(x|\bm{\lambda}\right) & \triangleq\mu_{n,\mathbf{Y}}\left(\left.\frac{\sum_{i=1}^{n}\lambda_{i}Y_{i}}{\sum_{i=1}^{n}Y_{i}}\le x\right|\bm{\lambda}\right)\label{eq:prob_Y_def}\\
 & =\mu_{n,\mathbf{Y}}\left(\left.\sum_{i=1}^{n}\left(\lambda_{i}-x\right)Y_{i}\le0\right|\bm{\lambda}\right).\nonumber \end{align}
 Then \[
\Pr\left(\left.\underset{k}{\min}\frac{\sum_{i=1}^{n}\lambda_{i}Y_{i}}{\sum_{i=1}^{n}Y_{i}}\le x\right|\bm{\lambda}\right)=1-\left(1-\mu_{n,\mathbf{Y}}\left(x|\bm{\lambda}\right)\right)^{2^{R_{\mathrm{fb}}}}.\]
 Thus\[
\mathrm{E}_{\mathcal{B}_{\mathrm{rand}}}\left[\underset{k}{\min}\frac{\sum_{i=1}^{n}\lambda_{i}\left|z_{k,i}\right|^{2}}{\sum_{i=1}^{n}\lambda_{i}\left|z_{k,i}\right|^{2}}\right]=\lambda_{\min,\bm{\lambda}}+\int_{\lambda_{\min,\bm{\lambda}}}^{\lambda_{\max,\bm{\lambda}}}\left(1-\mu_{n,\mathbf{Y}}\left(x|\bm{\lambda}\right)\right)^{2^{R_{\mathrm{fb}}}}dx\]
 and \begin{equation}
c_{\min,n,\mathrm{rand}}=\frac{\min\left(n,m\right)}{n}\mathrm{E}_{\bm{\lambda}}\left[\lambda_{\min,\bm{\lambda}}+\int_{\lambda_{\min,\bm{\lambda}}}^{\lambda_{\max,\bm{\lambda}}}\left(1-\mu_{n,\mathbf{Y}}\left(x|\bm{\lambda}\right)\right)^{2^{R_{\mathrm{fb}}}}dx\right].\label{eq:c_rand_formula}\end{equation}

In order to prove the bounds in (\ref{eq:c_rand_lb}) and (\ref{eq:c_rand_ub}),
we need the large deviations of $\mu_{n,\mathbf{Y}}\left(x|\bm{\lambda}\right)$
in Theorem \ref{thm:LD-Y}. Specifically, as $n,m\rightarrow\infty$
with $\frac{m}{n}\rightarrow\frac{1}{\beta}$, for $\forall x\in\left(\lambda_{t}^{-},\bar{\lambda}\right)$\begin{equation}
\underset{\left(n,m\right)\rightarrow\infty}{\lim}\frac{1}{n}\log\mu_{n,\mathbf{Y}}\left(\left.x\right|\bm{\lambda}\right)=-\psi_{x}^{*}\left(0\right)\label{eq:LD_Y_1}\end{equation}
 almost surely in $\bm{\lambda}$.

\subsubsection{Proof of the Lower Bound}

$ $

We prove the lower bound in (\ref{eq:c_rand_lb}). Take an $\epsilon>0$
small enough such that $\lambda_{t}^{-}<x_{r}^{-}-\epsilon$. Since
$\psi_{x_{r}^{-}-\epsilon}^{*}\left(0\right)>\psi_{x_{r}^{-}}^{*}\left(0\right)$
(Proposition \ref{pro:Psi^Star-x}(4)), there exists a $\delta_{\epsilon}>0$
s.t. $\psi_{x_{r}^{-}-\epsilon}^{*}\left(0\right)>\psi_{x_{r}^{-}}^{*}\left(0\right)+2\delta_{\epsilon}$
and $\lambda_{t}^{-}+\delta_{\epsilon}<x_{r}^{-}-\epsilon<\lambda^{+}-\delta_{\epsilon}$.
Define\begin{align*}
A_{n,\bm{\lambda}} & \triangleq\left\{ \bm{\lambda}:\;\left|\frac{1}{n}\log\mu_{n,\mathbf{Y}}\left(\left.x_{r}^{-}-\epsilon\right|\bm{\lambda}\right)+\psi_{x_{r}^{-}-\epsilon}^{*}\left(0\right)\right|<\delta_{\epsilon},\right.\\
 & \quad\quad\quad\quad\quad\left.\left|\lambda_{\min,\bm{\lambda}}-\lambda_{t}^{-}\right|<\delta_{\epsilon},\;\left|\lambda_{\max,\bm{\lambda}}-\lambda^{+}\right|<\delta_{\epsilon}\right\} .\end{align*}
 According to the large deviation principle in (\ref{eq:LD_Y_1})
and the almost sure convergence of $\lambda_{\min,n}$ and $\lambda_{\max,n}$
(Proposition \ref{prop:Asymptotic-Lambda-Max}), $\underset{\left(n,m\right)\rightarrow\infty}{\lim}\mu_{n,\bm{\lambda}}\left(A_{n,\bm{\lambda}}\right)=1$.
Note that on the set $A_{n,\bm{\lambda}}$\[
\mu_{n,\mathbf{Y}}\left(\left.x_{r}^{-}-\epsilon\right|\bm{\lambda}\right)\le e^{-n\left(\psi_{x_{r}^{-}-\epsilon}^{*}\left(0\right)-\delta_{\epsilon}\right)}\le e^{-n\left(\psi_{x_{r}^{-}}^{*}\left(0\right)+\delta_{\epsilon}\right)}.\]
 When $n$ is sufficiently large, on the set $A_{n,\bm{\lambda}}$
\begin{align*}
 & \left(1-\mu_{n,\mathbf{Y}}\left(\left.x_{r}^{-}-\epsilon\right|\bm{\lambda}\right)\right)^{2^{R_{\mathrm{fb}}}}\\
 & \ge\exp\left\{ 2^{R_{\mathrm{fb}}}\cdot\log\left(1-e^{-n\left(\psi_{x_{r}^{-}}^{*}\left(0\right)+\delta_{\epsilon}\right)}\right)\right\} \\
 & =\exp\left\{ e^{n\left(r\log2+o\left(1\right)\right)}\cdot\left[-e^{-n\left(\psi_{x_{r}^{-}}^{*}\left(0\right)+\delta_{\epsilon}\right)}\left(1+o\left(1\right)\right)\right]\right\} \\
 & =\exp\left\{ -e^{-n\left(\delta_{\epsilon}-o\left(1\right)\right)}\left(1+o\left(1\right)\right)\right\} \\
 & \ge1-\delta_{\epsilon}.\end{align*}
 Therefore, when $n$ is large enough,\begin{align*}
 & \mathrm{E}_{\bm{\lambda}}\left[\int_{\lambda_{\min,\bm{\lambda}}}^{\lambda_{\max,\bm{\lambda}}}\left(1-\mu_{n,\mathbf{Y}}\left(\left.x\right|\bm{\lambda}\right)\right)^{2^{R_{\mathrm{fb}}}}dx\right]\\
 & \ge\mathrm{E}_{\bm{\lambda}}\left[\int_{\lambda_{t}^{-}+\delta_{\epsilon}}^{x_{r}^{-}-\epsilon}\left(1-\mu_{n,\mathbf{Y}}\left(\left.x\right|\bm{\lambda}\right)\right)^{2^{R_{\mathrm{fb}}}}dx,\; A_{n,\bm{\lambda}}\right]\\
 & \ge\mathrm{E}_{\bm{\lambda}}\left[\int_{\lambda_{t}^{-}+\delta_{\epsilon}}^{x_{r}^{-}-\epsilon}\left(1-\mu_{n,\mathbf{Y}}\left(\left.x_{r}^{-}-\epsilon\right|\bm{\lambda}\right)\right)^{2^{R_{\mathrm{fb}}}}dx,\; A_{n,\bm{\lambda}}\right]\\
 & \ge\mathrm{E}_{\bm{\lambda}}\left[\int_{\lambda_{t}^{-}+\delta_{\epsilon}}^{x_{r}^{-}-\epsilon}\left(1-\delta_{\epsilon}\right),\; A_{n,\bm{\lambda}}\right]\\
 & \ge\left(1-\delta_{\epsilon}\right)^{2}\left(x_{r}^{-}-\epsilon-\lambda_{t}^{-}-\delta_{\epsilon}\right)\end{align*}
 where the last inequality follows from the fact that $\mu_{n,\bm{\lambda}}\left(A_{n,\bm{\lambda}}\right)\ge1-\delta_{\epsilon}$
for sufficiently large $n$. Decrease $\delta_{\epsilon}>0$ to zero
and then let $\epsilon>0$ approach zero. We have \[
\mathrm{E}_{\bm{\lambda}}\left[\int_{\lambda_{\min,\bm{\lambda}}}^{\lambda_{\max,\bm{\lambda}}}\left(1-\mu_{n,\mathbf{Y}}\left(\left.x\right|\bm{\lambda}\right)\right)^{2^{R_{\mathrm{fb}}}}dx\right]\ge x_{r}^{-}-\lambda_{t}^{-}.\]
 Substitute it into (\ref{eq:c_rand_formula}) and note that $\mathrm{E}_{\bm{\lambda}}\left[\lambda_{\min,\bm{\lambda}}\right]\rightarrow\lambda_{t}^{-}$
(Proposition \ref{prop:Asymptotic-Lambda-Max}). The lower bound (\ref{eq:c_rand_lb})
is proved.

\subsubsection{Proof of the Upper Bound}

$ $

Now we prove the upper bound in (\ref{eq:c_rand_ub}).

Take an $\epsilon>0$ small enough such that $x_{r}^{-}+\epsilon<\bar{\lambda}$.
Since $\psi_{x_{r}^{-}}^{*}\left(0\right)>\psi_{x_{r}^{-}+\epsilon}^{*}\left(0\right)$
(Proposition \ref{pro:Psi^Star-x}(4)), there exists a $\delta_{\epsilon}>0$
s.t. $\psi_{x_{r}^{-}}^{*}\left(0\right)>\psi_{x_{r}^{-}+\epsilon}^{*}\left(0\right)+2\delta_{\epsilon}$
and $\lambda_{t}^{-}+\delta_{\epsilon}<x_{r}^{-}+\epsilon<\lambda^{+}-\delta_{\epsilon}$.
Define\begin{align*}
B_{n,\bm{\lambda}} & \triangleq\left\{ \bm{\lambda}:\;\left|\frac{1}{n}\log\mu_{n,\mathbf{Y}}\left(\left.x_{r}^{-}+\epsilon\right|\bm{\lambda}\right)+\psi_{x_{r}^{-}+\epsilon}^{*}\left(0\right)\right|<\delta_{\epsilon},\right.\\
 & \quad\quad\quad\quad\quad\left.\left|\lambda_{\min,\bm{\lambda}}-\lambda_{t}^{-}\right|<\delta_{\epsilon},\;\left|\lambda_{\max,\bm{\lambda}}-\lambda^{+}\right|<\delta_{\epsilon}\right\} .\end{align*}
 Then $\underset{\left(n,m\right)\rightarrow\infty}{\lim}\mu_{n,\bm{\lambda}}\left(B_{n,\bm{\lambda}}\right)=1$.
Note that on the set $B_{n,\bm{\lambda}}$\[
\mu_{n,\mathbf{Y}}\left(\left.x_{r}^{-}+\epsilon\right|\bm{\lambda}\right)\ge e^{-n\left(\psi_{x_{r}^{-}+\epsilon}^{*}\left(0\right)+\delta_{\epsilon}\right)}\ge e^{-n\left(\psi_{x_{r}^{-}}^{*}\left(0\right)-\delta_{\epsilon}\right)}.\]
 When $n$ is sufficiently large, on the set $B_{n,\bm{\lambda}}$\begin{align*}
 & \left(1-\mu_{n,\mathbf{Y}}\left(\left.x_{r}^{-}+\epsilon\right|\bm{\lambda}\right)\right)^{2^{R_{\mathrm{fb}}}}\\
 & \le\exp\left\{ 2^{R_{\mathrm{fb}}}\log\left(1-e^{-n\left(\psi_{x_{r}^{-}}^{*}\left(0\right)-\delta_{\epsilon}\right)}\right)\right\} \\
 & =\exp\left\{ e^{n\left(r\log2+o\left(1\right)\right)}\cdot\left[-e^{-n\left(\psi_{x_{r}^{-}}^{*}\left(0\right)-\delta_{\epsilon}\right)}\left(1+o\left(1\right)\right)\right]\right\} \\
 & =\exp\left\{ -e^{n\left(\delta_{\epsilon}+o\left(1\right)\right)}\left(1+o\left(1\right)\right)\right\} \\
 & \le\delta_{\epsilon}.\end{align*}
 Note that\begin{align}
 & \mathrm{E}_{\bm{\lambda}}\left[\int_{\lambda_{\min,\bm{\lambda}}}^{\lambda_{\max,\bm{\lambda}}}\left(1-\mu_{n,\mathbf{Y}}\left(\left.x\right|\bm{\lambda}\right)\right)^{2^{R_{\mathrm{fb}}}}dx\right]\nonumber \\
 & \le\mathrm{E}_{\bm{\lambda}}\left[\int_{\lambda_{\min,\bm{\lambda}}}^{\lambda_{\max,\bm{\lambda}}}\left(1-\mu_{n,\mathbf{Y}}\left(\left.x\right|\bm{\lambda}\right)\right)^{2^{R_{\mathrm{fb}}}}dx,\; B_{n,\bm{\lambda}}\right]+\mathrm{E}_{\bm{\lambda}}\left[\int_{\lambda_{\min,\bm{\lambda}}}^{\lambda_{\max,\bm{\lambda}}}1\cdot dx,\; B_{n,\bm{\lambda}}^{c}\right].\label{eq:c_rand-ub-split}\end{align}
 The first term is upper bounded by \begin{align*}
 & \mathrm{E}_{\bm{\lambda}}\left[\int_{\lambda_{\min,\bm{\lambda}}}^{\lambda_{\max,\bm{\lambda}}}\left(1-\mu_{n,\mathbf{Y}}\left(\left.x\right|\bm{\lambda}\right)\right)^{2^{R_{\mathrm{fb}}}}dx,\; B_{n,\bm{\lambda}}\right]\\
 & \le\mathrm{E}_{\bm{\lambda}}\left[\int_{\lambda_{t}^{-}-\delta_{\epsilon}}^{x_{r}^{-}+\epsilon}1dx+\int_{x_{r}^{-}+\epsilon}^{\lambda^{+}+\delta_{\epsilon}}\left(1-\mu_{n,\mathbf{Y}}\left(\left.x\right|\bm{\lambda}\right)\right)^{2^{R_{\mathrm{fb}}}}dx,\; B_{n,\bm{\lambda}}\right]\\
 & \le\left(x_{r}^{-}+\epsilon-\lambda_{t}^{-}+\delta_{\epsilon}\right)\mu_{n,\bm{\lambda}}\left(B_{n,\bm{\lambda}}\right)\\
 & \quad+\mathrm{E}_{\bm{\lambda}}\left[\int_{x_{r}^{-}+\epsilon}^{\lambda^{+}+\delta_{\epsilon}}\left(1-\mu_{n,\mathbf{Y}}\left(\left.x_{r}^{-}+\epsilon\right|\bm{\lambda}\right)\right)^{2^{R_{\mathrm{fb}}}}dx,\; B_{n,\bm{\lambda}}\right]\\
 & \le\left(x_{r}^{-}+\epsilon-\lambda_{t}^{-}+\delta_{\epsilon}\right)\mu_{n,\bm{\lambda}}\left(B_{n,\bm{\lambda}}\right)+\left(\lambda^{+}+\delta_{\epsilon}-x_{r}^{-}-\epsilon\right)\delta_{\epsilon}\cdot\mu_{n,\bm{\lambda}}\left(B_{n,\bm{\lambda}}\right)\\
 & \le\left[\left(x_{r}^{-}+\epsilon-\lambda_{t}^{-}+\delta_{\epsilon}\right)+\left(\lambda^{+}+\delta_{\epsilon}-x_{r}^{-}-\epsilon\right)\delta_{\epsilon}\right]\left(1-\delta_{\epsilon}\right)\end{align*}
 when $n$ is sufficiently large. The second term in (\ref{eq:c_rand-ub-split})
can be upper bounded by \begin{align*}
 & \mathrm{E}_{\bm{\lambda}}\left[\int_{\lambda_{\min,\bm{\lambda}}}^{\lambda_{\max,\bm{\lambda}}}1\cdot dx,\; B_{n,\bm{\lambda}}^{c}\right]\\
 & =\mathrm{E}_{\bm{\lambda}}\left[\lambda_{\max,\bm{\lambda}},\; B_{n,\bm{\lambda}}^{c}\right]-\mathrm{E}_{\bm{\lambda}}\left[\lambda_{\min,\bm{\lambda}},\; B_{n,\bm{\lambda}}^{c}\right]\\
 & \le\mathrm{E}_{\bm{\lambda}}\left[\lambda_{\max,\bm{\lambda}},\; B_{n,\bm{\lambda}}^{c}\right]+\mathrm{E}_{\bm{\lambda}}\left[\lambda_{\min,\bm{\lambda}},\; B_{n,\bm{\lambda}}^{c}\right]\\
 & \le2\delta_{\epsilon},\end{align*}
 for sufficiently large $n$, where the last inequality is implied
by Proposition \ref{prop:Asymptotic-Lambda-Max}. Let $\delta_{\epsilon}\downarrow0$
and then $\epsilon\downarrow0$. \[
\underset{\left(n,m\right)\rightarrow\infty}{\lim}\mathrm{E}_{\bm{\lambda}}\left[\int_{\lambda_{\min,\bm{\lambda}}}^{\lambda_{\max,\bm{\lambda}}}\left(1-\mu_{n,\mathbf{Y}}\left(\left.x\right|\bm{\lambda}\right)\right)^{2^{R_{\mathrm{fb}}}}dx\right]\le x_{r}^{-}-\lambda_{t}^{-},\]
 and therefore the upper bound (\ref{eq:c_rand_ub}) is proved.

\subsection{\label{sub:bounds-arbitrary-codebook}Uniform Bounds for Arbitrary
Codebooks}

$ $

Here we prove Theorem \ref{thm:asymptotic-bds-all-codebooks} for
which the following fact is important. Let $\mathbf{U}\in\mathcal{U}_{n\times n}$
be isotropically distributed, then for any given $\mathbf{v}\in\mathcal{U}_{n\times1}$
and $\bm{\lambda}$, $\mathbf{U}^{\dagger}\mathbf{v}\in\mathcal{U}_{n\times1}$
is isotropically distributed and \[
\mu_{n,\mathbf{U}}\left(\left.\mathbf{v}^{\dagger}\mathbf{U}\mathbf{\Lambda}\mathbf{U}^{\dagger}\mathbf{v}\le x\right|\mathbf{v},\bm{\lambda}\right)=\mu_{n,\mathbf{Y}}\left(\left.x\right|\bm{\lambda}\right),\]
 where $\mu_{n,\mathbf{Y}}\left(\left.x\right|\bm{\lambda}\right)$
is defined in (\ref{eq:prob_Y_def}). Furthermore, as $\lambda_{\min,\bm{\lambda}}<\lambda_{\max,\bm{\lambda}}$,
$\mu_{n,\mathbf{Y}}\left(\left.\lambda_{\min,\bm{\lambda}}\right|\bm{\lambda}\right)=0$,
$\mu_{n,\mathbf{Y}}\left(\left.\lambda_{\max,\bm{\lambda}}\right|\bm{\lambda}\right)=1$,
and there exists a unique $x_{p}\in\left(\lambda_{\min,\bm{\lambda}},\lambda_{\max,\bm{\lambda}}\right)$
such that $\mu_{n,\mathbf{Y}}\left(\left.x_{p}\right|\bm{\lambda}\right)=p$.

Recall the definitions in (\ref{eq:c_min_n_arbitrary_code}) and (\ref{eq:c_max_n_arbitrary_code}).
For any given $n\in\mathbb{N}$, singular value vector $\bm{\lambda}$
and codebook $\mathcal{B}$, the following lemma provides lower and
upper bounds on $c_{\min,n,\bm{\lambda},\mathcal{B}}$ and $c_{\max,n,\bm{\lambda},\mathcal{B}}$.

\begin{lemma}
\label{lem:lb-finite-size-codebook}Let $\bm{\lambda}$ be such that
$\lambda_{\min,\bm{\lambda}}<\lambda_{\max,\bm{\lambda}}$. Then \[
\underset{\mathcal{B}:\;\left|\mathcal{B}\right|=2^{R_{\mathrm{fb}}}}{\inf}\; c_{\min,n,\bm{\lambda},\mathcal{B}}\ge2^{R_{\mathrm{fb}}}\int_{\lambda_{\min,\bm{\lambda}}}^{x_{n,\bm{\lambda}}^{-}}x\cdot d\mu_{n,\mathbf{Y}}\left(\left.x\right|\bm{\lambda}\right)\]
 and \[
\underset{\mathcal{B}:\;\left|\mathcal{B}\right|=2^{R_{\mathrm{fb}}}}{\sup}\; c_{\max,n,\bm{\lambda},\mathcal{B}}\le2^{R_{\mathrm{fb}}}\int_{x_{n,\bm{\lambda}}^{+}}^{\lambda_{\max,\bm{\lambda}}}x\cdot d\mu_{n,\mathbf{Y}}\left(\left.x\right|\bm{\lambda}\right)\]
 where $\mu_{n,\mathbf{Y}}\left(\left.x_{n,\bm{\lambda}}^{-}\right|\bm{\lambda}\right)=2^{-R_{\mathrm{fb}}}$
and $\mu_{n,\mathbf{Y}}\left(\left.x_{n,\bm{\lambda}}^{+}\right|\bm{\lambda}\right)=1-2^{-R_{\mathrm{fb}}}$. 
\end{lemma}
\begin{proof}
We give the details behind the lower bound on $c_{\min,n,\bm{\lambda},\mathcal{B}}$
omitting those for $c_{\max,n,\bm{\lambda},\mathcal{B}}$. For any
given $\mathcal{B}$ such that $\left|\mathcal{B}\right|=2^{R_{\mathrm{fb}}}$,
\begin{align*}
 & \mu_{n,\mathbf{U}}\left(\left.\underset{\mathbf{v}\in\mathcal{B}}{\min}\mathbf{v}^{\dagger}\mathbf{U}\mathbf{\Lambda}\mathbf{U}^{\dagger}\mathbf{v}\le x\right|\bm{\lambda},\mathcal{B}\right)\\
 & =\mu\left(\left.\cup_{k=1}^{2^{R_{\mathrm{fb}}}}\left\{ \mathbf{U}\in\mathcal{U}_{n\times n}:\;\mathbf{v}_{k}^{\dagger}\mathbf{U}\mathbf{\Lambda}\mathbf{U}^{\dagger}\mathbf{v}_{k}=\underset{\mathbf{v}\in\mathcal{B}}{\min}\mathbf{v}^{\dagger}\mathbf{U}\mathbf{\Lambda}\mathbf{U}^{\dagger}\mathbf{v}\le x\right\} \right|\bm{\lambda},\mathcal{B}\right)\\
 & \le\sum_{k=1}^{2^{R_{\mathrm{fb}}}}\mu\left(\left.\left\{ \mathbf{U}\in\mathcal{U}_{n\times n}:\;\mathbf{v}_{k}^{\dagger}\mathbf{U}\mathbf{\Lambda}\mathbf{U}^{\dagger}\mathbf{v}_{k}=\underset{\mathbf{v}\in\mathcal{B}}{\min}\mathbf{v}^{\dagger}\mathbf{U}\mathbf{\Lambda}\mathbf{U}^{\dagger}\mathbf{v}\le x\right\} \right|\bm{\lambda},\mathcal{B}\right)\\
 & \le\sum_{k=1}^{2^{R_{\mathrm{fb}}}}\mu_{n,\mathbf{U}}\left\{ \left.\mathbf{U}\in\mathcal{U}_{n\times n}:\;\mathbf{v}_{k}^{\dagger}\mathbf{U}\mathbf{\Lambda}\mathbf{U}^{\dagger}\mathbf{v}_{k}\le x\right|\bm{\lambda},\mathbf{v}_{k}\right\} \\
 & =2^{R_{\mathrm{fb}}}\mu_{n,\mathbf{Y}}\left(\left.x\right|\bm{\lambda}\right).\end{align*}
 Thus, \begin{align*}
c_{\min,n,\bm{\lambda},\mathcal{B}} & =\int_{\lambda_{\min,\bm{\lambda}}}^{\lambda_{\max,\bm{\lambda}}}x\cdot d\mu_{n,\mathbf{U}}\left(\left.x\right|\bm{\lambda},\mathcal{B}\right)\\
 & =\lambda_{\max,\bm{\lambda}}-\int_{\lambda_{\min,\bm{\lambda}}}^{\lambda_{\max,\bm{\lambda}}}\mu_{n,\mathbf{U}}\left(\left.x\right|\bm{\lambda},\mathcal{B}\right)dx\\
 & \ge\lambda_{\max,\bm{\lambda}}-\int_{\lambda_{\min,\bm{\lambda}}}^{\lambda_{\max,\bm{\lambda}}}\min\left(2^{R_{\mathrm{fb}}}\mu_{n,\mathbf{Y}}\left(\left.x\right|\bm{\lambda}\right),1\right)dx\\
 & =\int_{\lambda_{\min,\bm{\lambda}}}^{\lambda_{\max,\bm{\lambda}}}x\cdot d\min\left(2^{R_{\mathrm{fb}}}\mu_{n,\mathbf{Y}}\left(\left.x\right|\bm{\lambda}\right),1\right)\\
 & =2^{R_{\mathrm{fb}}}\int_{\lambda_{\min,\bm{\lambda}}}^{x_{n,\bm{\lambda}}^{-}}x\cdot d\mu_{n,\mathbf{Y}}\left(\left.x\right|\bm{\lambda}\right).\end{align*}
 The proof is finished. 
\end{proof}

The next lemma shows that $x_{n,\bm{\lambda}}^{\pm}$ converge $\bm{\lambda}$-almost
surely to the advertised constants.

\begin{lemma}
\label{lem:Convergence-x_n}As $n,m,R_{\mathrm{fb}}\rightarrow\infty$
linearly with $\frac{m}{n}\rightarrow\beta\in\mathbb{R}^{+}$ and
$\frac{R_{\mathrm{fb}}}{n}\rightarrow r\in\mathbb{R}^{+}$, $\underset{\left(n,m,R_{\mathrm{fb}}\right)\rightarrow\infty}{\lim}x_{n,\bm{\lambda}}^{-}=x_{r}^{-}$
and $\underset{\left(n,m,R_{\mathrm{fb}}\right)\rightarrow\infty}{\lim}x_{n,\bm{\lambda}}^{+}=x_{r}^{+}$
almost surely in $\bm{\lambda}$. 
\end{lemma}
\begin{proof}
Take the case of $x_{n,\bm{\lambda}}^{-}$, that for $x_{n,\bm{\lambda}}^{+}$
being much the same. Note that $\psi_{x}^{*}\left(0\right)$ monotone
decreases as $x$ increases in $\left(-\frac{1}{x-\lambda_{t}^{-}},0\right)$
(Proposition \ref{pro:Psi^Star-x}(4)). For $\forall\epsilon>0$ small
enough such that $\lambda_{t}^{-}<x_{r}^{-}-\epsilon<x_{r}^{-}+\epsilon<\bar{\lambda}$,
there exists a $\delta_{\epsilon}>0$ such that $\psi_{x_{r}^{-}-\epsilon}^{*}\left(0\right)-\delta_{\epsilon}>r\log2=\psi_{x_{r}^{-}}^{*}\left(0\right)>\psi_{x_{r}^{-}+\epsilon}^{*}\left(0\right)+\delta_{\epsilon}$.
According to the large deviation principle in (\ref{eq:LD_Y_1}),
\[
\underset{\left(n,m\right)\rightarrow\infty}{\lim}\frac{1}{n}\log\mu_{n,\mathbf{Y}}\left(\left.x_{r}^{-}-\epsilon\right|\bm{\lambda}\right)<-\left(r\log2+\delta_{\epsilon}\right)\]
 and \[
\underset{\left(n,m\right)\rightarrow\infty}{\lim}\frac{1}{n}\log\mu_{n,\mathbf{Y}}\left(\left.x_{r}^{-}+\epsilon\right|\bm{\lambda}\right)>-\left(r\log2-\delta_{\epsilon}\right)\]
 almost surely in $\bm{\lambda}$. By the definition of $x_{n,\bm{\lambda}}^{-}$,\[
\underset{\left(n,m,R_{\mathrm{fb}}\right)\rightarrow\infty}{\lim}\frac{1}{n}\log\mu_{n,\mathbf{v}}\left(\left.x_{n,\bm{\lambda}}^{-}\right|\bm{\lambda}\right)==\underset{\left(n,m,R_{\mathrm{fb}}\right)\rightarrow\infty}{\lim}\frac{1}{n}\log2^{-R_{\mathrm{fb}}}=-r\log2.\]
 Therefore, $x_{r}^{-}-\epsilon<\underset{\left(n,m,R_{\mathrm{fb}}\right)\rightarrow\infty}{\lim}x_{n,\bm{\lambda}}^{-}<x_{r}^{-}+\epsilon$
almost surely. To finish, let $\epsilon\downarrow0$. 
\end{proof}

Now we are ready to prove Theorem \ref{thm:asymptotic-bds-all-codebooks}.

\begin{proof}
{[}Proof of Theorem \ref{thm:asymptotic-bds-all-codebooks}] Once
again, we only give the details for $c_{\min}$. 
\begin{enumerate}
\item Take an $\epsilon>0$ small enough such that $\lambda_{t}^{-}<x_{r}^{-}-2\epsilon$.
Since $\psi_{x_{r}^{-}-\epsilon}^{*}\left(0\right)>\psi_{x_{r}^{-}}^{*}\left(0\right)$,
$\exists\delta_{\epsilon}>0$ s.t. $\psi_{x_{r}^{-}-\epsilon}^{*}\left(0\right)>\psi_{x_{r}^{-}}^{*}\left(0\right)+2\delta_{\epsilon}$.
Define a set \begin{align*}
A_{n,\bm{\lambda}} & =\left\{ \bm{\lambda}:\;\left|\lambda_{\min,\bm{\lambda}}-\lambda_{t}^{-}\right|\le\epsilon,\;\left|x_{n,\bm{\lambda}}^{-}-x_{r}^{-}\right|\le\epsilon\right.\\
 & \quad\quad\left.\left|\frac{1}{n}\log\mu_{n,\mathbf{Y}}\left(\left.x_{r}^{-}-\epsilon\right|\bm{\lambda}\right)+\psi_{x_{r}^{-}-\epsilon}^{*}\left(0\right)\right|\le\delta_{\epsilon}\right\} .\end{align*}
 According to Proposition \ref{prop:Asymptotic-Lambda-Max}, Lemma
\ref{lem:Convergence-x_n} and (\ref{eq:LD_Y_1}), $\mu_{n,\bm{\lambda}}\left(A_{n,\bm{\lambda}}\right)\overset{\left(n,m\right)\rightarrow\infty}{\longrightarrow}1$.
On the set $A_{n,\bm{\lambda}}$, when $n$ is sufficiently large,
\begin{align*}
 & 2^{R_{\mathrm{fb}}}\int_{\lambda_{\min,\bm{\lambda}}}^{x_{n,\bm{\lambda}}^{-}}x\cdot d\mu_{n,\mathbf{Y}}\left(\left.x\right|\bm{\lambda}\right)\\
 & =x_{n,\bm{\lambda}}^{-}-\int_{\lambda_{\min,\bm{\lambda}}}^{x_{n,\bm{\lambda}}^{-}}\min\left(2^{R_{\mathrm{fb}}}\mu_{n,\mathbf{Y}}\left(\left.x\right|\bm{\lambda}\right),1\right)dx\\
 & \ge\left(x_{r}^{-}-\epsilon\right)-2^{R_{\mathrm{fb}}}\mu_{n,\mathbf{Y}}\left(\left.x_{r}^{-}-\epsilon\right|\bm{\lambda}\right)\int_{\lambda_{\min,\bm{\lambda}}}^{x_{r}^{-}-\epsilon}dx-\int_{x_{r}^{-}-\epsilon}^{x_{r}^{-}+\epsilon}1dx\\
 & \ge x_{r}^{-}-3\epsilon-e^{n\left(r\log2+o\left(1\right)\right)}e^{-n\left(\psi_{x_{r}^{-}-\epsilon}^{*}\left(0\right)-\delta_{\epsilon}\right)}\left(x_{r}^{-}-\epsilon-\lambda_{\min,\bm{\lambda}}\right)\\
 & \ge x_{r}^{-}-3\epsilon-e^{-n\left(\delta_{\epsilon}+o\left(1\right)\right)}\left(x_{r}^{-}-\epsilon-\lambda_{t}^{-}+\epsilon\right)\\
 & \ge x_{r}^{-}-4\epsilon.\end{align*}
 Now take $\epsilon\downarrow0$ yielding part (1) of Theorem \ref{thm:asymptotic-bds-all-codebooks}.
\item For any $\epsilon>0$, define\[
B_{n,\bm{\lambda}}\triangleq\left\{ \bm{\lambda}:\;\underset{\mathcal{B}:\;\left|\mathcal{B}\right|=2^{R_{\mathrm{fb}}}}{\inf}\; c_{\min,n,\bm{\lambda},\mathcal{B}}\ge x_{r}^{-}-\epsilon\right\} .\]
 From Part (1), $\mu_{n,\bm{\lambda}}\left(B_{n,\bm{\lambda}}\right)\overset{\left(n,m\right)\rightarrow\infty}{\longrightarrow}1$.
On the set $B_{n,\bm{\lambda}}$, for sufficiently large $n$, \begin{align*}
 & \underset{\mathcal{B}:\;\left|\mathcal{B}\right|=2^{R_{\mathrm{fb}}}}{\inf}\;\mathrm{E}_{\bm{\lambda}}\left[\mathrm{E}_{\mathbf{U}}\left[\underset{\mathbf{v}\in\mathcal{B}}{\min}\mathbf{v}^{\dagger}\mathbf{U}\mathbf{\Lambda}\mathbf{U}^{\dagger}\mathbf{v}\right]\right]\\
 & \ge\mathrm{E}_{\bm{\lambda}}\left[\underset{\mathcal{B}:\;\left|\mathcal{B}\right|=2^{R_{\mathrm{fb}}}}{\inf}\mathrm{E}_{\mathbf{U}}\left[\underset{\mathbf{v}\in\mathcal{B}}{\min}\mathbf{v}^{\dagger}\mathbf{U}\mathbf{\Lambda}\mathbf{U}^{\dagger}\mathbf{v}\right]\right]\\
 & \ge\mathrm{E}_{\bm{\lambda}}\left[\underset{\mathcal{B}:\;\left|\mathcal{B}\right|=2^{R_{\mathrm{fb}}}}{\inf}c_{\min,n,\bm{\lambda},\mathcal{B}},\; B_{n,\bm{\lambda}}\right]\\
 & \ge\left(x_{r}^{-}-\epsilon\right)\left(1-\epsilon\right).\end{align*}
 Again, $\epsilon$ can now be taken to zero to complete the proof. 
\end{enumerate}
\end{proof}

\subsection{\label{sub:Optimality-Random-Codes}Asymptotic Optimality of the
Random Codebooks}

$ $

At last we come to the proof of Theorem \ref{thm:Optimality-Random-Codes}.
As before, it is enough to focus on the $x_{r}^{-}$ case.

While the proof of Theorem \ref{thm:Average-Performance-Random-Codes}
rests on the probability measure $\mu_{n,\mathcal{B}_{\mathrm{rand}}}\left(\left.\cdot\right|\bm{\lambda},\mathbf{U}\right)$,
we now require the measure $\mu_{n,\bm{\lambda},\mathbf{U}}\left(\left.\cdot\right|\mathcal{B}_{\mathrm{rand}}\right)$.
These two measures are connected by the joint measure $\mu_{n,\mathcal{B}_{\mathrm{rand}},\bm{\lambda},\mathbf{U}}$:
for any measurable set $A\subset\left\{ \mathcal{B}_{\mathrm{rand}}\right\} \times\left\{ \bm{\lambda}\right\} \times\left\{ \mathbf{U}\right\} $,
\begin{align*}
\mu_{n,\mathcal{B}_{\mathrm{rand}},\bm{\lambda},\mathbf{U}} & =\mathrm{E}_{\bm{\lambda},\mathbf{U}}\left[\mu_{n,\mathcal{B}_{\mathrm{rand}}}\left(\left.A\right|\bm{\lambda},\mathbf{U}\right)\right]\\
 & =\mathrm{E}_{\mathcal{B}_{\mathrm{rand}}}\left[\mu_{n,\bm{\lambda},\mathbf{U}}\left(\left.A\right|\mathcal{B}_{\mathrm{rand}}\right)\right].\end{align*}

We first show that for any $\epsilon>0$, \begin{equation}
\underset{\left(n,m,R_{\mathrm{fb}}\right)\rightarrow\infty}{\lim}\;\mu_{n,\mathcal{B}_{\mathrm{rand}},\bm{\lambda},\mathbf{U}}\left(\underset{\mathbf{v}\in\mathcal{B}_{\mathrm{rand}}}{\min}\;\mathbf{v}^{\dagger}\mathbf{U}\mathbf{\Lambda}\mathbf{U}^{\dagger}\mathbf{v}\le x_{r}^{-}+\epsilon\right)=1.\label{eq:optimality-asymp-prob-behavior}\end{equation}
 Note that $\psi_{x_{r}^{-}+\epsilon}^{*}\left(0\right)<r\log2$.
There exists a $\delta_{\epsilon}>0$ s.t. $\psi_{x_{r}^{-}+\epsilon}^{*}\left(0\right)+2\delta_{\epsilon}<r\log2$.
Let \[
A_{n,\bm{\lambda}}=\left\{ \bm{\lambda}:\;\left|\frac{1}{n}\log\mu_{n,\mathbf{Y}}\left(\left.x_{r}^{-}+\epsilon\right|\bm{\lambda}\right)+\psi_{x_{r}^{-}+\epsilon}^{*}\left(0\right)\right|<\delta_{\epsilon}\right\} .\]
 Then $\mu_{n,\bm{\lambda}}\left(A_{n,\bm{\lambda}}\right)\overset{\left(n,m\right)\rightarrow\infty}{\longrightarrow}1$
by (\ref{eq:LD_Y_1}). Thus, as $n$ is large enough, \begin{align*}
 & \mu_{n,\mathcal{B}_{\mathrm{rand}},\bm{\lambda},\mathbf{U}}\left(\underset{\mathbf{v}\in\mathcal{B}_{\mathrm{rand}}}{\min}\;\mathbf{v}^{\dagger}\mathbf{\Lambda}\mathbf{v}\le x_{r}^{-}+\epsilon\right)\\
 & =\mathrm{E}_{\bm{\lambda},\mathbf{U}}\left[\mu_{n,\mathcal{B}_{\mathrm{rand}}}\left(\left.\underset{\mathbf{v}\in\mathcal{B}_{\mathrm{rand}}}{\min}\;\mathbf{v}^{\dagger}\mathbf{\Lambda}\mathbf{v}\le x_{r}^{-}+\epsilon\right|\bm{\lambda},\mathbf{U}\right)\right]\\
 & =\mathrm{E}_{\bm{\lambda},\mathbf{U}}\left[1-\left(1-\mu_{n,\mathbf{Y}}\left(\left.x_{r}^{-}+\epsilon\right|\bm{\lambda}\right)\right)^{2^{R_{\mathrm{fb}}}}\right]\\
 & \ge\left(1-\exp\left\{ -e^{n\left(r\log2+o\left(1\right)\right)}e^{-n\left(\psi_{x_{r}^{-}+\epsilon}^{*}\left(0\right)+\delta_{\epsilon}+o\left(1\right)\right)}\left(1+o\left(1\right)\right)\right\} \right)\mu_{n,\bm{\lambda}}\left(A_{n,\bm{\lambda}}\right)\\
 & \ge\left(1-\delta_{\epsilon}\right)\left(1-\delta_{\epsilon}\right).\end{align*}
 This is (\ref{eq:optimality-asymp-prob-behavior}) once $\delta_{\epsilon}\downarrow0$.

Next we have the following fact. For $\forall\epsilon>0$, let $\delta_{\epsilon}>0$
be such that $\left(\lambda^{+}+\epsilon\right)\delta_{\epsilon}<\frac{\epsilon}{4}$.
Define a set \[
B_{n}\triangleq\left\{ \mathcal{B}_{\mathrm{rand}}:\;\mu_{n,\bm{\lambda},\mathbf{U}}\left(\left.\underset{\mathbf{v}\in\mathcal{B}_{\mathrm{rand}}}{\min}\;\mathbf{v}^{\dagger}\mathbf{U}\mathbf{\Lambda}\mathbf{U}^{\dagger}\mathbf{v}\le x_{r}^{-}+\frac{\epsilon}{4}\right|\mathcal{B}_{\mathrm{rand}}\right)>1-\delta_{\epsilon}\right\} .\]
 Then \[
\underset{\left(n,m,R_{\mathrm{fb}}\right)\rightarrow\infty}{\lim}\;\mu_{n,\mathcal{B}_{\mathrm{rand}}}\left(B_{n}\right)=1.\]
 This fact can be proved by contradiction. If it were not true there
would exist a subsequence $n_{j}$ such that $\mu_{n_{j},\mathcal{B}_{\mathrm{rand}}}\left(B_{n_{j}}\right)<1-t$
for some $t>0$, and \begin{align*}
 & \underset{\left(n,m,R_{\mathrm{fb}}\right)\rightarrow\infty}{\lim}\;\mu_{n,\mathcal{B}_{\mathrm{rand}},\bm{\lambda},\mathbf{U}}\left(\underset{\mathbf{v}\in\mathcal{B}_{\mathrm{rand}}}{\min}\;\mathbf{v}^{\dagger}\mathbf{\Lambda}\mathbf{v}\le x_{r}^{-}+\epsilon\right)\\
 & \le\underset{\left(n,m,R_{\mathrm{fb}}\right)\rightarrow\infty}{\lim}1\cdot\mu_{n_{j},\mathcal{B}_{\mathrm{rand}}}\left(B_{n_{j}}\right)+\left(1-\delta_{\epsilon}\right)\mu_{n_{j},\mathcal{B}_{\mathrm{rand}}}\left(B_{n_{j}}^{c}\right)\\
 & <1,\end{align*}
 which contradicts (\ref{eq:optimality-asymp-prob-behavior}).

Now on the set $B_{n}$, if $n$ is large enough, \begin{align*}
 & \mathrm{E}_{\bm{\lambda},\mathbf{U}}\left[\left.\underset{\mathbf{v}\in\mathcal{B}_{\mathrm{rand}}}{\min}\;\mathbf{v}^{\dagger}\mathbf{U}\mathbf{\Lambda}\mathbf{U}^{\dagger}\mathbf{v}\right|\mathcal{B}_{\mathrm{rand}}\right]\\
 & \overset{\left(a\right)}{\le}\left(x_{r}^{-}+\frac{\epsilon}{4}\right)\mu_{n,\bm{\lambda},\mathbf{U}}\left(\left.\underset{\mathbf{v}\in\mathcal{B}_{\mathrm{rand}}}{\min}\;\mathbf{v}^{\dagger}\mathbf{U}\mathbf{\Lambda}\mathbf{U}^{\dagger}\mathbf{v}\le x_{r}^{-}+\frac{\epsilon}{4}\right|\mathcal{B}_{\mathrm{rand}}\right)\\
 & \quad+\mathrm{E}_{\bm{\lambda},\mathbf{U}}\left[\left.\lambda_{\max,\bm{\lambda}},\left|\lambda_{\max,\bm{\lambda}}-\lambda^{+}\right|\le\epsilon,\underset{\mathbf{v}\in\mathcal{B}_{\mathrm{rand}}}{\min}\;\mathbf{v}^{\dagger}\mathbf{U}\mathbf{\Lambda}\mathbf{U}^{\dagger}\mathbf{v}>x_{r}^{-}+\frac{\epsilon}{4}\right|\mathcal{B}_{\mathrm{rand}}\right]\\
 & \quad+\mathrm{E}_{\bm{\lambda},\mathbf{U}}\left[\left.\lambda_{\max,\bm{\lambda}},\left|\lambda_{\max,\bm{\lambda}}-\lambda^{+}\right|>\epsilon,\underset{\mathbf{v}\in\mathcal{B}_{\mathrm{rand}}}{\min}\;\mathbf{v}^{\dagger}\mathbf{U}\mathbf{\Lambda}\mathbf{U}^{\dagger}\mathbf{v}>x_{r}^{-}+\frac{\epsilon}{4}\right|\mathcal{B}_{\mathrm{rand}}\right]\\
 & \le\left(x_{r}^{-}+\frac{\epsilon}{4}\right)\cdot1+\left(\lambda^{+}+\epsilon\right)\mu_{n,\bm{\lambda},\mathbf{U}}\left(\left.\underset{\mathbf{v}\in\mathcal{B}_{\mathrm{rand}}}{\min}\;\mathbf{v}^{\dagger}\mathbf{U}\mathbf{\Lambda}\mathbf{U}^{\dagger}\mathbf{v}>x_{r}^{-}+\frac{\epsilon}{4}\right|\mathcal{B}_{\mathrm{rand}}\right)\\
 & \quad+\mathrm{E}_{\bm{\lambda},\mathbf{U}}\left[\left.\lambda_{\max,\bm{\lambda}},\left|\lambda_{\max,\bm{\lambda}}-\lambda^{+}\right|>\epsilon\right|\mathcal{B}_{\mathrm{rand}}\right]\\
 & \le\left(x_{r}^{-}+\frac{\epsilon}{4}\right)+\left(\lambda^{+}+\epsilon\right)\delta_{\epsilon}+\mathrm{E}_{\bm{\lambda}}\left[\lambda_{\max,\bm{\lambda}},\left|\lambda_{\max,\bm{\lambda}}-\lambda^{+}\right|>\epsilon\right]\\
 & \overset{\left(b\right)}{\le}\left(x_{r}^{-}+\frac{\epsilon}{4}\right)+\frac{\epsilon}{4}+\frac{\epsilon}{2}=x_{r}^{-}+\epsilon,\end{align*}
 where

$\left(a\right)$ follows from the fact that $\underset{\mathbf{v}\in\mathcal{B}_{\mathrm{rand}}}{\min}\;\mathbf{v}^{\dagger}\mathbf{U}\mathbf{\Lambda}\mathbf{U}^{\dagger}\mathbf{v}\le\lambda_{\max,\bm{\lambda}}$,
and

$\left(b\right)$ follows from the fact that $\mathrm{E}_{\bm{\lambda}}\left[\lambda_{\max,\bm{\lambda}},\left|\lambda_{\max,\bm{\lambda}}-\lambda^{+}\right|>\epsilon\right]\le\frac{\epsilon}{2}$
for sufficiently large $n$.

Therefore, Theorem \ref{thm:Optimality-Random-Codes} is proved.

\section{Simulations}

\begin{figure}
\subfigure[$\beta=1/2$]{\includegraphics[scale=0.5]{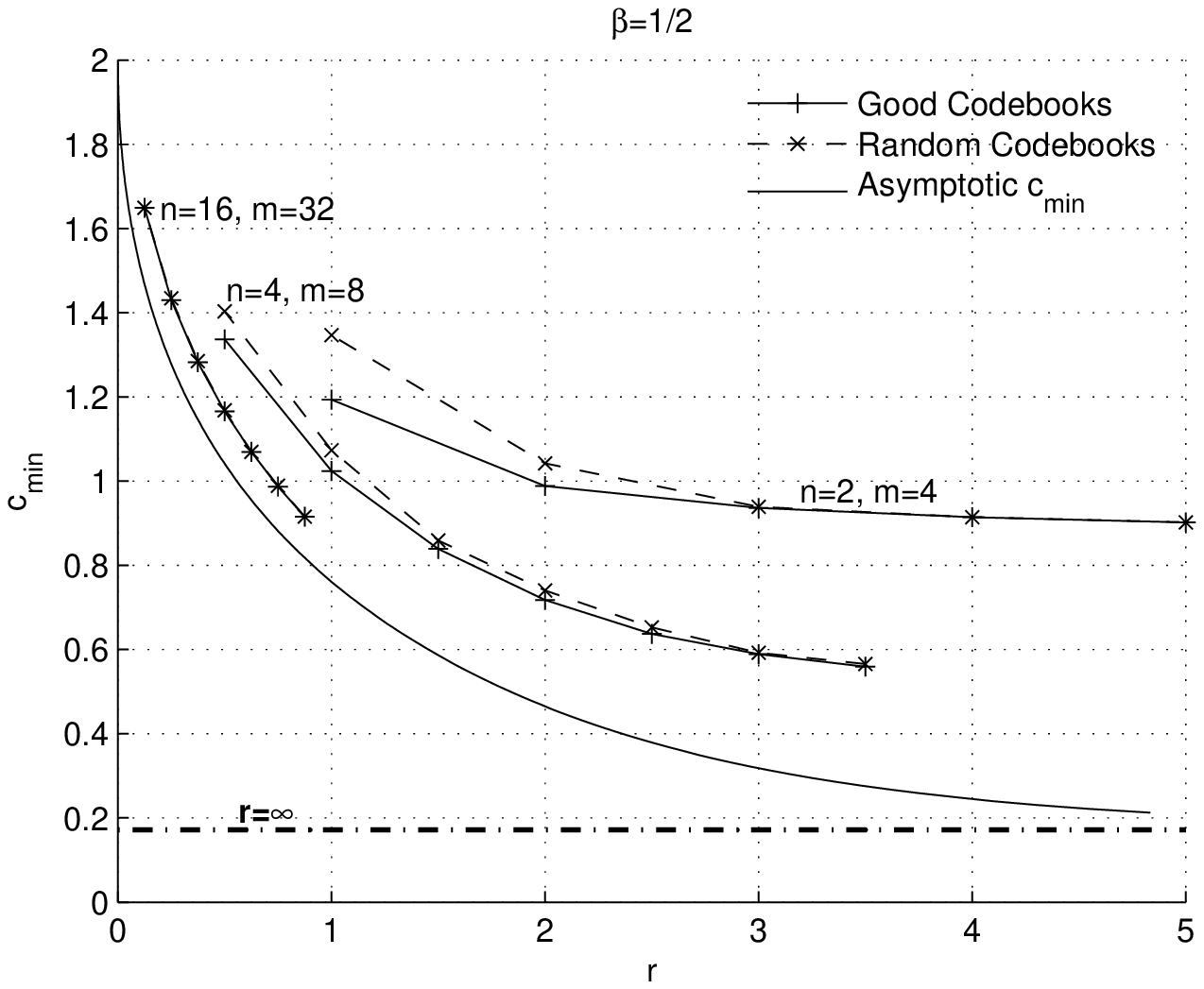}}\subfigure[$\beta=2$]{\includegraphics[scale=0.5]{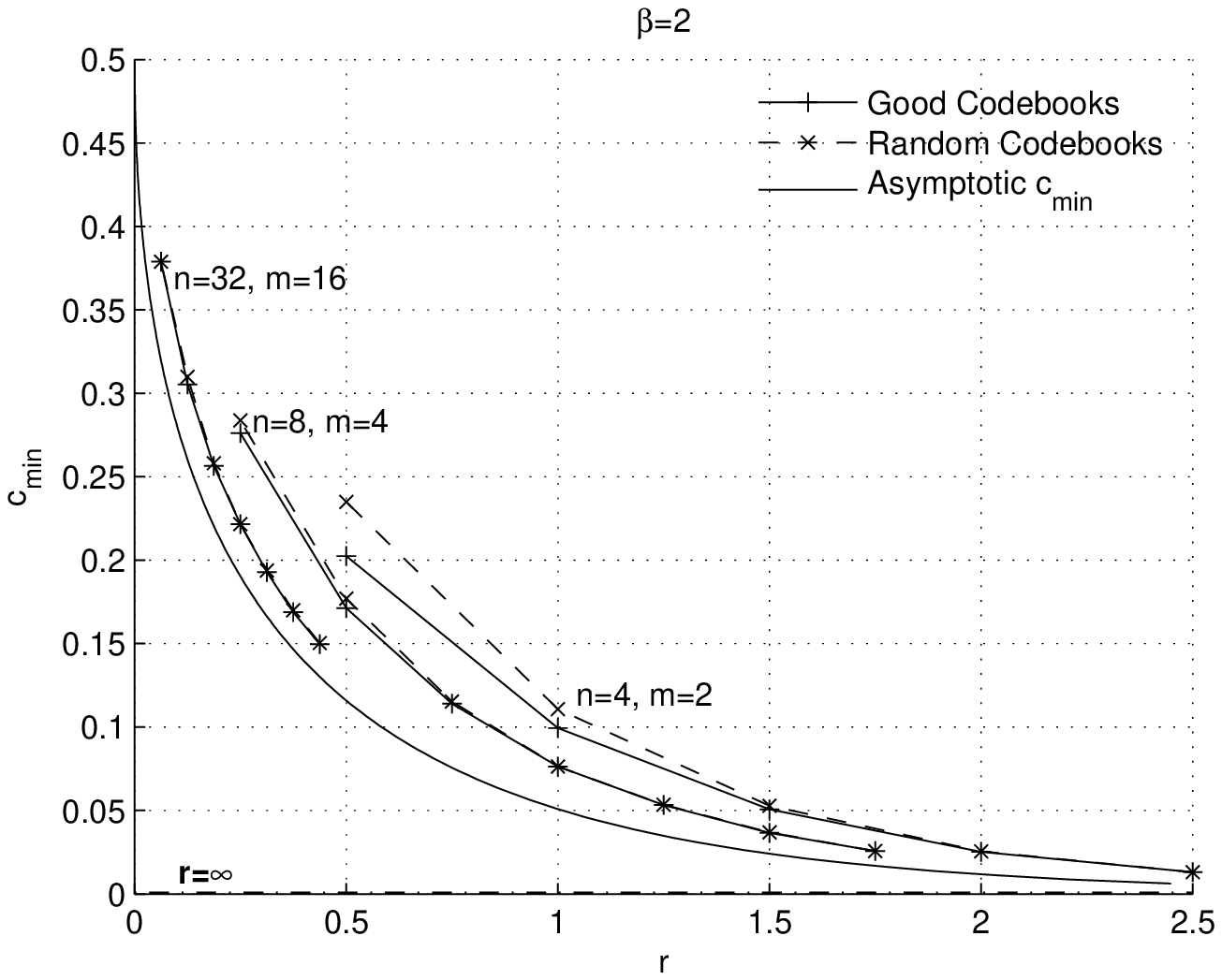}}

\caption{\label{fig:c_min_n}$c_{\min,n}$}
\end{figure}

\begin{figure}
\subfigure[$\beta=1/2$]{\includegraphics[clip,scale=0.5]{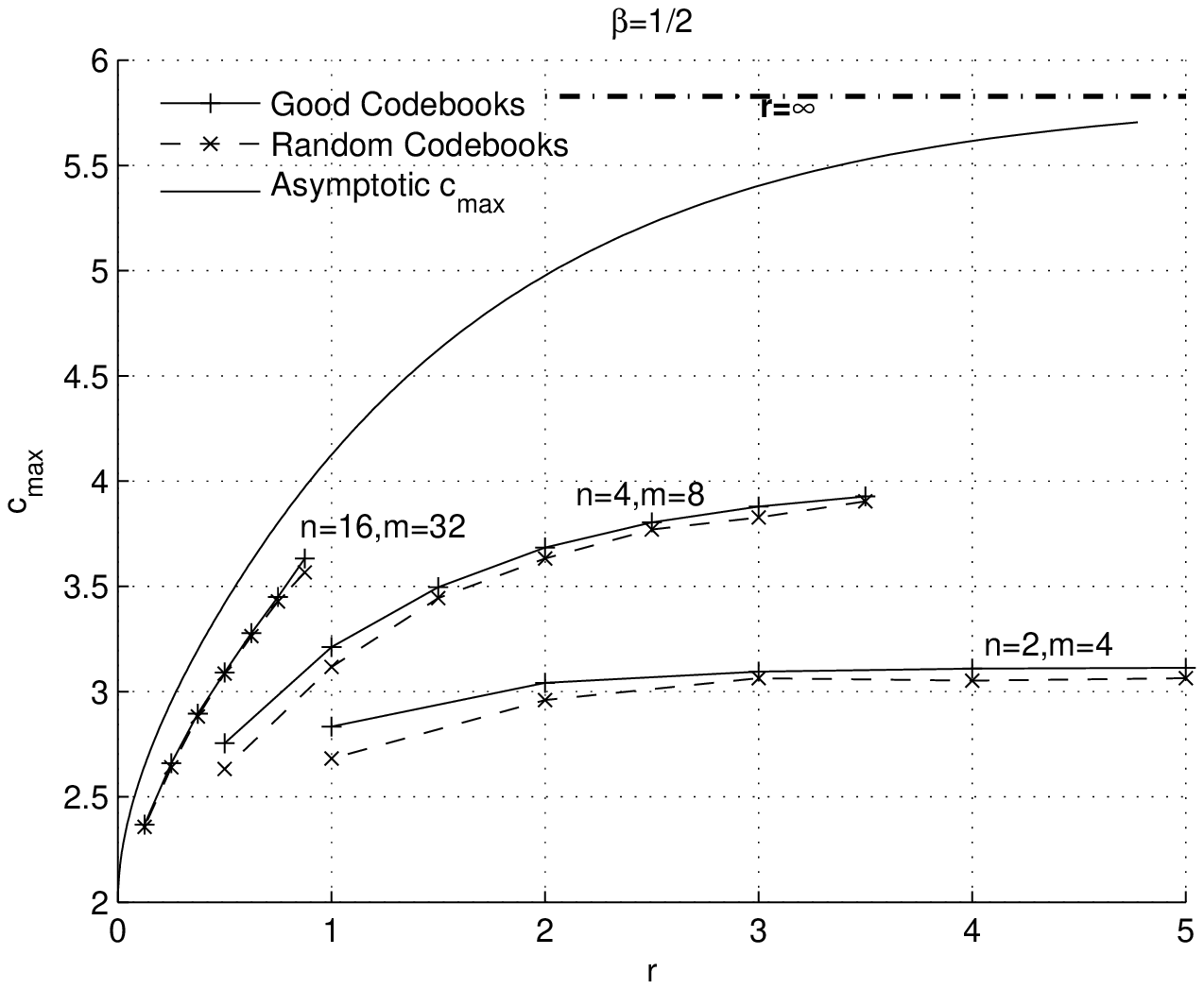}}\subfigure[$\beta=2$]{\includegraphics[clip,scale=0.5]{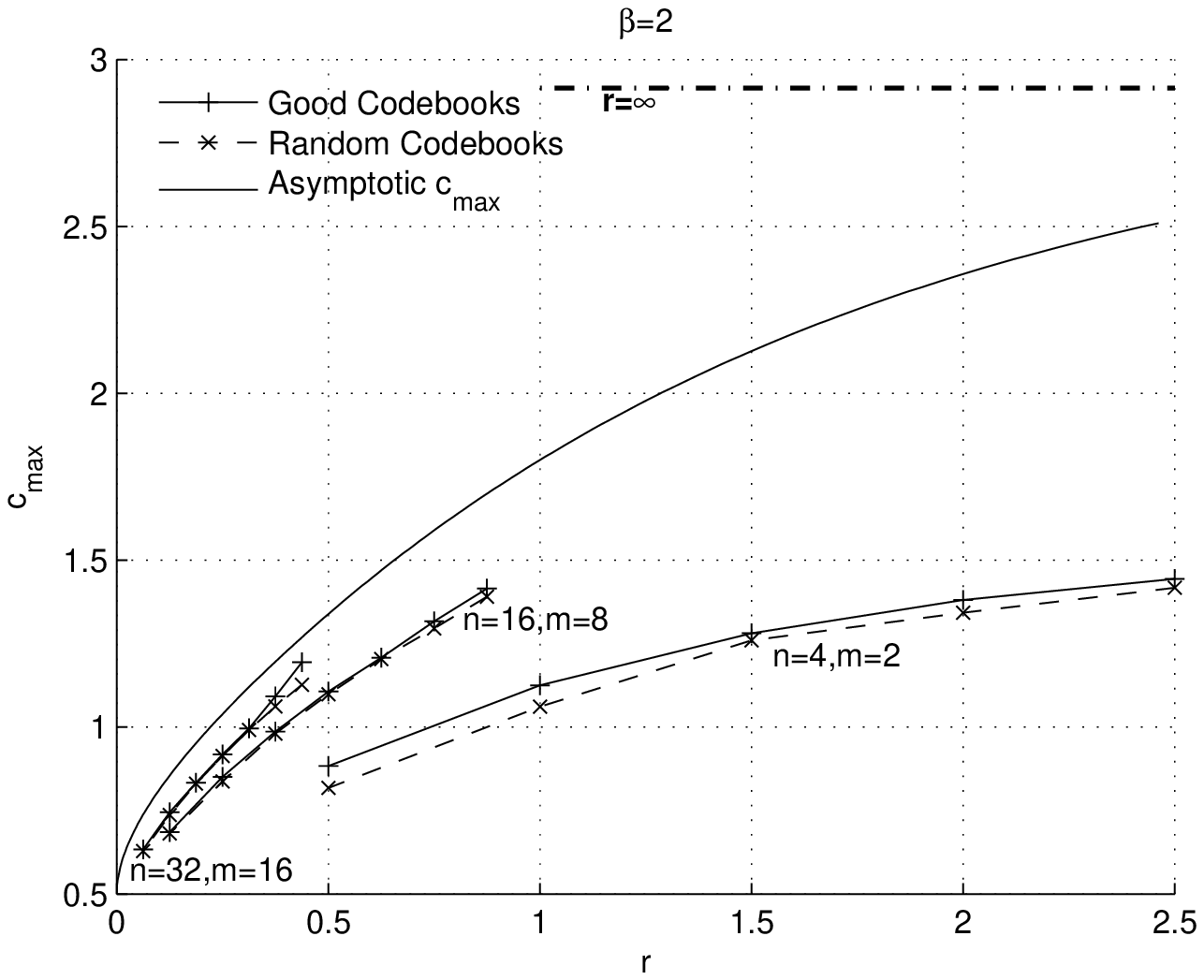}}

\caption{\label{fig:c_max_n}$c_{\max,n}$}
\end{figure}

Fig \ref{fig:c_min_n} and \ref{fig:c_max_n} give simulation results
for several CDMA systems and MIMO systems respectively. In both figures,
the $x$ axis is the normalized feedback rate $r=\frac{R_{\mathrm{fb}}}{n}$.
The $y$ axis in Fig \ref{fig:c_min_n} is the $c_{\min,n}$ and that
in Fig \ref{fig:c_max_n} is the $c_{\max,n}$. The dashed lines with
x markers are for random codebooks while the solid lines with plus
markers are for well designed codebooks, which are numerically generated
by the criterion of maximizing the minimum chordal distance of the
codebook. The solid lines without any markers are the asymptotic performance
by Corollary \ref{cor:main-result-computations}. Simulations show
that as $n,m,R_{\mathrm{fb}}$ increase linearly, the performance
($c_{\min,n}$ and $c_{\max,n}$) will get closer to the asymptotic
one. Although random codebooks are not optimal for finite dimensional
systems, as $n,m,R_{\mathrm{fb}}$ increase linearly, the difference
between random codebooks and well-designed codebooks decreases.

\section{Conclusion}

In this paper, we analyze the effect of finite rate feedback on CDMA
signature optimization and MIMO beamforming vector selection. The
main results are the exact asymptotic performance formulae. In addition,
we prove that random codebooks are asymptotically optimal not only
on average but also with probability one. The proofs rest on a large
deviation principle derived over a random matrix ensemble.

\appendix

\subsection{\label{sec:Properties-of-Rate-Functions}Properties of Rate Functions}

Let $Y$ be a non-negative random variable with probability measure
$d\mu_{y}=e^{-y}dy$ for $y\in\left[0,+\infty\right)$. Let $Y_{M}$
be a non-negative random variable with probability measure \begin{equation}
d\mu_{y,M}=\begin{cases}
e^{-y}\frac{dy}{\mu_{Y}\left[0,M\right]} & \mathrm{if}\; y\in\left[0,M\right]\\
0 & \mathrm{otherwise}\end{cases}.\label{eq:u_Y_M}\end{equation}
 Let $d\mu_{\lambda}$ be (\ref{eq:spectrum-pdf}). Define the moment
generating functions \begin{equation}
\psi_{x,M}\left(\alpha\right)\triangleq\int\log\mathrm{E}_{Y}\left[e^{\alpha\left(\lambda-x\right)Y},\left|Y\right|\le M\right]d\mu_{\lambda},\label{eq:mgf_M}\end{equation}
 \begin{equation}
\psi_{x}\left(\alpha\right)\triangleq\underset{M\rightarrow\infty}{\lim}\psi_{x,M}\left(\alpha\right),\label{eq:mgf_Y}\end{equation}
 and \begin{equation}
\psi_{x,Y_{M}}\left(\alpha\right)\triangleq\int\log\mathrm{E}_{Y_{M}}\left[e^{\alpha\left(\lambda-x\right)Y_{M}}\right]d\mu_{\lambda}.\label{eq:mgf_Y_M}\end{equation}
 Clearly, $\psi_{x,Y_{M}}\left(\alpha\right)=\psi_{x,M}\left(\alpha\right)-\log\mu_{Y}\left[0,M\right].$
Proposition \ref{pro:Psi-Properties}(3) shows that $\psi_{x}\left(\alpha\right)$
has the form (\ref{eq:moment-generate-fn-def}). Furthermore, for
$\forall t\in\mathbb{R}$, define the rate functions

\begin{equation}
\psi_{x}^{*}\left(t\right)\triangleq\underset{\alpha\in\mathbb{R}}{\sup}\left[\alpha t-\psi_{x}\left(\alpha\right)\right],\label{eq:rate_fn_Y}\end{equation}
 and \begin{equation}
\psi_{x,Y_{M}}^{*}\left(t\right)\triangleq\underset{\alpha\in\mathbb{R}}{\sup}\left[\alpha t-\psi_{x,Y_{M}}\left(\alpha\right)\right].\label{eq:rate_fn_Y_M}\end{equation}
 In the following, we shall discuss the properties of these moment
generating functions and rate functions.

\begin{prop}
\label{pro:Psi-Properties}(Properties of $\psi\left(\cdot\right)$'s) 
\begin{enumerate}
\item For $\forall\alpha\in\mathbb{R}$, $\log\mathrm{E}_{Y}\left[e^{\alpha\left(\lambda-x\right)Y},\left|Y\right|\le M\right]$
is a Lipschitz function on any compact set of $\lambda\in\mathbb{R}$. 
\item $\psi_{x,M}\left(\alpha\right)$ is monotonically increasing with
$M$, and \begin{align*}
\psi_{x}\left(\alpha\right) & \triangleq\underset{M\rightarrow\infty}{\lim}\psi_{x,M}\left(\alpha\right)\\
 & =\begin{cases}
-\int\log\left(1+\alpha\left(x-\lambda\right)\right)d\mu_{\lambda} & \mathrm{if}\;\alpha\in\left[-\frac{1}{x-\lambda_{t}^{-}},\frac{1}{\lambda^{+}-x}\right]\\
+\infty & \mathrm{otherwise}\end{cases}.\end{align*}

\item $\psi_{x,M}\left(\alpha\right)$ and $\psi_{x,Y_{M}}\left(\alpha\right)$
are strictly convex functions of $\alpha\in\mathbb{R}$, and $\psi_{x}\left(\alpha\right)$
is strictly convex on $\alpha\in\left(-\frac{1}{x-\lambda_{t}^{-}},\frac{1}{\lambda^{+}-x}\right)$. 
\item Let $x\in\left(\lambda_{t}^{-},\lambda^{+}\right)$. For $\forall t<\bar{\lambda}-x$,
if $M$ is large enough, there exists an $\alpha\in\left(-\infty,0\right)$
such that $\psi_{x,M}^{\prime}\left(\alpha\right)=\psi_{x,Y_{M}}^{\prime}\left(\alpha\right)=t$.
Similarly, for $\forall t>\bar{\lambda}-x$, if $M$ is large enough,
there exists an $\alpha\in\left(0,+\infty\right)$ such that $\psi_{x,M}^{\prime}\left(\alpha\right)=\psi_{x,Y_{M}}^{\prime}\left(\alpha\right)=t$. 
\end{enumerate}
\end{prop}
\begin{proof}
$ $ 
\begin{enumerate}
\item First note with the restriction $\{|Y|\le M\}$, both $\log\mathrm{E}_{Y}\left[e^{\alpha\left(\lambda-x\right)Y},\left|Y\right|\le M\right]$
and $\psi_{x,M}\left(\alpha\right)$ are well defined. Let $f_{M}\left(\lambda\right)=\log\int_{0}^{M}e^{\alpha\left(\lambda-x\right)y}e^{-y}dy$.
Then \[
f_{M}^{\prime}\left(\lambda\right)=\frac{\int_{0}^{M}\alpha ye^{\alpha\left(\lambda-x\right)y}e^{-y}dy}{\int_{0}^{M}e^{\alpha\left(\lambda-x\right)y}e^{-y}dy}.\]
 Since both $e^{\alpha\left(\lambda-x\right)y-y}$ and $\alpha ye^{\alpha\left(\lambda-x\right)y-y}$
are continuous functions of $\lambda$ and $y$ on the compact set
$A_{\lambda}\times\left[0,M\right]$, and there exist positive constants
$a>0$ and $b>0$ such that \[
e^{\alpha\left(\lambda-x\right)y}e^{-y}\ge a,\ \ \mbox{ and }\ \ \left|\alpha ye^{\alpha\left(\lambda-x\right)y-y}\right|\le b\]
 on that set. Thus, $\left|f_{M}^{\prime}\left(\lambda\right)\right|\le\frac{b}{a}<\infty$
and so $f_{M}\left(\lambda\right)$ is Lipschitz on $A_{\lambda}$. 
\item The monotonicity of $\psi_{x,M}\left(\alpha\right)$ is obvious, as
is the identification of the limit $\psi_{x,M}\left(\alpha\right)$
presented in Part (3). 
\item The convexity of logarithmic moment generating functions is a standard
fact, see for example Chapter 2 of \cite{Dembo_book92_Large_Deviation}. 
\item It is clear that $\psi_{x,M}^{\prime}\left(\alpha\right)=\psi_{x,Y_{M}}^{\prime}\left(\alpha\right)$,
and we need only calculate the former.

To simplify the notation, denote $z=1-\alpha\left(\lambda-x\right)$.
Then\[
\psi_{x,M}\left(\alpha\right)=\int\log\left(\frac{1-e^{-Mz}}{z}\right)d\mu_{\lambda}\]
 and \begin{align}
\psi_{x,M}^{\prime}\left(\alpha\right) & =\int\frac{1-\left(1+Mz\right)e^{-Mz}}{z\left(1-e^{-Mz}\right)}\left(\lambda-x\right)\cdot d\mu_{\lambda}.\label{eq:psi_M_derivative_integral}\end{align}

If $\alpha=0$, then $z\equiv1$ and \[
\psi_{x,M}^{\prime}\left(0\right)=\frac{1-\left(1+M\right)e^{-M}}{1-e^{-M}}\left(\bar{\lambda}-x\right).\]
 It is clear that \[
\psi_{x,M}^{\prime}\left(0\right)\overset{M\rightarrow+\infty}{\longrightarrow}\bar{\lambda}-x.\]

Now we evaluate $\psi_{x,M}^{\prime}\left(\pm\infty\right)$. Note
that \[
z=1-\alpha\left(\lambda-x\right)\overset{\alpha\rightarrow-\infty}{\longrightarrow}\begin{cases}
-\infty & \mathrm{on}\;\lambda\in\left[\lambda_{t}^{-},x\right)\\
+\infty & \mathrm{on}\;\lambda\in\left(x,\lambda^{+}\right]\end{cases}.\]
 We evaluate the integrand in (\ref{eq:psi_M_derivative_integral})
and obtain that \[
\frac{1-\left(1+Mz\right)e^{-Mz}}{z\left(1-e^{-Mz}\right)}\left(\lambda-x\right)\overset{\alpha\rightarrow-\infty}{\longrightarrow}\begin{cases}
M\left(\lambda-x\right) & \mathrm{on}\;\lambda\in\left[\lambda_{t}^{-},x\right)\\
0 & \mathrm{on}\;\lambda\in\left(x,\lambda^{+}\right]\end{cases},\]
 and \[
\frac{1-\left(1+Mz\right)e^{-Mz}}{z\left(1-e^{-Mz}\right)}\left(\lambda-x\right)\overset{\alpha\rightarrow+\infty}{\longrightarrow}\begin{cases}
0 & \mathrm{on}\;\lambda\in\left[\lambda_{t}^{-},x\right)\\
M\left(\lambda-x\right) & \mathrm{on}\;\lambda\in\left(x,\lambda^{+}\right]\end{cases}.\]
 Therefore, \[
\psi_{x,M}^{\prime}\left(-\infty\right)\overset{M\rightarrow+\infty}{\longrightarrow}-\infty,\]
 and \[
\psi_{x,M}^{\prime}\left(+\infty\right)\overset{M\rightarrow+\infty}{\longrightarrow}+\infty,\]
 which prove Part (5). 

\end{enumerate}
\end{proof}
\begin{prop}
\label{pro:Psi^Star-M}(Properties of $\psi_{x,Y_{M}}^{*}\left(\cdot\right)$) 
\begin{enumerate}
\item $\psi_{x,Y_{M}}^{*}\left(t\right)\ge0$. 
\item For $\forall x\in\left(\lambda_{t}^{-},\lambda^{+}\right)$ and $\forall t\in\mathbb{R}$,
if $M$ is large enough, there exists a $\gamma\in\mathbb{R}$ such
that $\psi_{x,Y_{M}}^{\prime}\left(\gamma\right)=t$ and $\psi_{x,Y_{M}}^{*}\left(t\right)=\gamma t-\psi_{x,Y_{M}}\left(\gamma\right)$.
More specifically, $\gamma<0$ when $t<\bar{\lambda}-x$, and $\gamma>0$
when $t>\bar{\lambda}-x$ . 
\item Consider an $x\in\left(\lambda_{t}^{-},\lambda^{+}\right)$ and $\forall t_{1},t_{2}\in\mathbb{R}$.
For a sufficiently large $M$, let $\psi_{x,Y_{M}}^{*}\left(t_{1}\right)$
and $\psi_{x,Y_{M}}^{*}\left(t_{2}\right)$ be achieved at $\alpha\in\mathbb{R}$
and $\beta\in\mathbb{R}$ respectively. Then \[
\psi_{x,Y_{M}}^{*}\left(t_{2}\right)-\psi_{x,Y_{M}}^{*}\left(t_{1}\right)\le\beta\left(t_{2}-t_{1}\right),\]
 where the equality holds if and only if $t_{1}=t_{2}$. 
\end{enumerate}
\end{prop}
\begin{proof}
$ $
\begin{enumerate}
\item It is from the fact that $\left.\alpha t-\psi_{x,Y_{M}}\left(\alpha\right)\right|_{\alpha=0}=0$. 
\item This part follows directly from Proposition \ref{pro:Psi-Properties}(5)
and the strict convexity of $\psi_{x,Y_{M}}^{*}\left(t\right)$ (Proposition
\ref{pro:Psi-Properties}(4)). 
\item Let $\alpha\in\mathbb{R}$ and $\gamma\in\mathbb{R}$ be such that
$\psi_{x,Y_{M}}^{\prime}\left(\alpha\right)=t_{1}$ and $\psi_{x,Y_{M}}^{\prime}\left(\gamma\right)=t_{2}$.
Clearly, $\psi_{x,Y_{M}}^{*}\left(t_{1}\right)=\alpha t_{1}-\psi_{x,Y_{M}}\left(\alpha\right)$
and $\psi_{x,Y_{M}}^{*}\left(t_{2}\right)=\gamma t_{2}-\psi_{x,Y_{M}}\left(\gamma\right)$.
If $t_{1}\neq t_{2}$, then $\alpha\neq\gamma$ and \begin{align*}
\psi_{x,Y_{M}}^{*}\left(t_{2}\right)-\psi_{x,Y_{M}}^{*}\left(t_{1}\right) & =\gamma t_{2}-\psi_{x,Y_{M}}\left(\gamma\right)-\left(\alpha t_{1}-\psi_{x,Y_{M}}\left(\alpha\right)\right)\\
 & =\gamma\left(t_{2}-t_{1}\right)+t_{1}\left(\gamma-\alpha\right)-\left[\psi_{x,Y_{M}}\left(\gamma\right)-\psi_{x,Y_{M}}\left(\alpha\right)\right]\\
 & \overset{\left(a\right)}{=}\gamma\left(t_{2}-t_{1}\right)+t_{1}\left(\gamma-\alpha\right)-\psi_{x,Y_{M}}^{\prime}\left(\xi\right)\left(\gamma-\alpha\right)\\
 & =\gamma\left(t_{2}-t_{1}\right)+\left(t_{1}-\psi_{x,Y_{M}}^{\prime}\left(\xi\right)\right)\left(\gamma-\alpha\right)\\
 & \overset{\left(b\right)}{<}\gamma\left(t_{2}-t_{1}\right),\end{align*}
 where $\left(a\right)$ is from the mean value theorem for some $\xi\in\left(\min\left(\alpha,\gamma\right),\max\left(\alpha,\gamma\right)\right)$,
and $\left(b\right)$ follows from the strict convexity of $\psi_{x,Y_{M}}\left(\alpha\right)$.
\end{enumerate}
\end{proof}
\begin{prop}
\label{pro:Psi^Star-x}(Properties of $\psi_{x}^{*}\left(0\right)$) 
\begin{enumerate}
\item $\psi_{x}^{*}\left(t\right)\ge0$. 
\item Let $x\in\left(\lambda_{t}^{-},\lambda^{+}\right)$. For $\forall t<\bar{\lambda}-x$,
\[
\psi_{x}^{*}\left(t\right)=\underset{\alpha\in\left(-\frac{1}{x-\lambda_{t}^{-}},0\right)}{\sup}\;\alpha t-\psi_{x}\left(\alpha\right).\]
 For $\forall t>\bar{\lambda}-x$, \[
\psi_{x}^{*}\left(t\right)=\underset{\alpha\in\left(0,\frac{1}{\lambda^{+}-x}\right)}{\sup}\;\alpha t-\psi_{x}\left(\alpha\right).\]

\item $\psi_{\bar{\lambda}}^{*}\left(0\right)=0$. 
\item $\psi_{x}^{*}\left(0\right)$ monotonically decreases on $x\in\left(\lambda_{t}^{-},\bar{\lambda}\right)$
and monotonically increases on $x\in\left(\bar{\lambda},\lambda^{+}\right)$. 
\item As $x\downarrow\lambda_{t}^{-}$ or $x\uparrow\lambda^{+}$, $\psi_{x}^{*}\left(0\right)\rightarrow+\infty$. 
\item For $\forall r\in\mathbb{R}^{+}$, there are unique $x_{r}^{-}\in\left(\lambda_{t}^{-},\bar{\lambda}\right)$
and $x_{r}^{+}\in\left(\bar{\lambda},\lambda^{+}\right)$ such that
$r\log2=\psi_{x_{r}^{-}}^{*}\left(0\right)=\psi_{x_{r}^{+}}^{*}\left(0\right)$. 
\end{enumerate}
\end{prop}
\begin{proof}
\begin{enumerate}
\item It is from the fact that $\left.\alpha t-\psi_{x}\left(\alpha\right)\right|_{\alpha=0}=0$. 
\item We only prove it for the case that $t<\bar{\lambda}-x$, as $t>\bar{\lambda}-x$
is the dual case.

It is clear that \[
\psi_{x}^{*}\left(t\right)=\underset{\alpha\in\left[-\frac{1}{x-\lambda_{t}^{-}},\frac{1}{\lambda^{+}-x}\right]}{\sup}\;\alpha t-\psi_{x}\left(\alpha\right)\]
 for $\psi_{x}\left(\alpha\right)=+\infty$ if $\alpha\notin\left[-\frac{1}{x-\lambda_{t}^{-}},\frac{1}{\lambda^{+}-x}\right]$.
Now for any $\alpha>0$, \begin{align*}
 & \alpha t-\int\log\mathrm{E}_{Y}\left[e^{\alpha\left(\lambda-x\right)Y}\right]d\mu_{\lambda}\\
 & \le\alpha t-\int\mathrm{E}_{Y}\left[\log e^{\alpha\left(\lambda-x\right)Y}\right]d\mu_{\lambda}\\
 & =\alpha\left(t-\left(\bar{\lambda}-x\right)\right)\\
 & <0.\end{align*}
 Note that $\psi_{x}^{*}\left(t\right)\ge0$, the $\sup$ is on $\alpha\in\left[-\frac{1}{x-\lambda_{t}^{-}},0\right]$.
Since $\log\left(1+\alpha\left(x-\lambda\right)\right)$ is continuous
on $\alpha\in\left(-\frac{1}{x-\lambda_{t}^{-}},0\right)$, it is
sufficient to have the sup on $\alpha\in\left(-\frac{1}{x-\lambda_{t}^{-}},0\right)$.

\item $\psi_{\bar{\lambda}}^{*}\left(0\right)\ge0$ by Part (1). However,
$\psi_{\bar{\lambda}}\left(\alpha\right)\ge-\log\left(1+\alpha\left(\bar{\lambda}-\bar{\lambda}\right)\right)=0$
and $\psi_{x}^{*}\left(t\right)=-\underset{\alpha\in\mathbb{R}}{\inf}\;\psi_{x}\left(\alpha\right)\le0$.
We have that $\psi_{\bar{\lambda}}^{*}\left(0\right)=0$. 
\item If $\lambda_{t}^{-}<y<x<\bar{\lambda}$, \begin{align*}
\psi_{y}^{*}\left(0\right) & =\underset{\alpha\in\left(-\frac{1}{y-\lambda_{t}^{-}},0\right)}{\sup}\;\int_{\lambda_{t}^{-}}^{\lambda^{+}}\log\left(1+\alpha\left(y-\lambda\right)\right)d\mu_{\lambda}\\
 & \overset{\left(a\right)}{\ge}\underset{\alpha\in\left(-\frac{1}{x-\lambda_{t}^{-}},0\right)}{\sup}\;\int_{\lambda_{t}^{-}}^{\lambda^{+}}\log\left(1+\alpha\left(y-\lambda\right)\right)d\mu_{\lambda}\\
 & \overset{\left(b\right)}{>}\underset{\alpha\in\left(-\frac{1}{x-\lambda_{t}^{-}},0\right)}{\sup}\;\int_{\lambda_{t}^{-}}^{\lambda^{+}}\log\left(1+\alpha\left(x-\lambda\right)\right)d\mu_{\lambda}\\
 & =\psi_{x}^{*}\left(0\right),\end{align*}
 where $\left(a\right)$ follows from shrinking the range of $\alpha$,
and $\left(b\right)$ is from the facts that $y-\lambda<x-\lambda$
and $\alpha<0$.

Similarly, if $\bar{\lambda}<x<y<\lambda^{+}$, $\psi_{y}^{*}\left(0\right)>\psi_{x}^{*}\left(0\right)$.

\item In order to prove that $x\downarrow\lambda_{t}^{-}$ implies $\psi_{x}^{*}\left(0\right)\uparrow\infty$,
let $x_{n}\downarrow\lambda_{t}^{-}$ and $\alpha_{n}=-\frac{1}{2\left(x_{n}-\lambda_{t}^{-}\right)}\in\left(-\frac{1}{x_{n}-\lambda_{t}^{-}},0\right)$.
Then $\psi_{x_{n}}\left(\alpha_{n}\right)$ is well defined for all
$n$.\[
\psi_{x_{n}}^{*}\left(0\right)\ge\psi_{x_{n}}\left(\alpha_{n}\right)=\int\log\left(\frac{x_{n}-\lambda_{t}^{-}+\lambda-\lambda_{t}^{-}}{2\left(x_{n}-\lambda_{t}^{-}\right)}\right)d\mu_{\lambda}.\]
 We shall show that $\psi_{x_{n}}^{*}\left(\alpha_{n}\right)\uparrow\infty$.

When $\beta\le1$ $\left(\lambda_{t}^{-}=\lambda^{-}\ge0\right)$,
\begin{align*}
\psi_{x_{n}}\left(\alpha_{n}\right) & =\int\log\left(x_{n}-\lambda_{t}^{-}+\lambda-\lambda_{t}^{-}\right)d\mu_{\lambda}-\log2-\log\left(x_{n}-\lambda_{t}^{-}\right).\end{align*}
 Since for $\epsilon>0$, $\left|\int_{0}^{\epsilon}x^{a+1}\log\left(x\right)dx\right|<\infty$
for $\forall a>-1$, and we have $\left|\int\log\left(\lambda-\lambda^{-}\right)d\mu_{\lambda}\right|<\infty$,
it holds \[
\psi_{x_{n}}\left(\alpha_{n}\right)\ge\int\log\left(\lambda-\lambda_{t}^{-}\right)d\mu_{\lambda}-\log\left(x_{n}-\lambda_{t}^{-}\right)-const.\overset{n\rightarrow\infty}{\rightarrow}+\infty.\]

When $\beta>1$ $\left(\lambda_{t}^{-}=0,\;\lambda^{-}>0\right)$,
\begin{align*}
\psi_{x_{n}}\left(\alpha_{n}\right) & =\frac{\tau-1}{\tau}\log\frac{1}{2}+\frac{1}{\tau}\int_{\lambda^{-}}^{\lambda^{+}}\log\left(\frac{x_{n}+\lambda}{2x_{n}}\right)\frac{\left(\lambda-\lambda^{-}\right)^{1/2}\left(\lambda^{+}-\lambda\right)^{\frac{1}{2}}}{2\pi\lambda}d\lambda\\
 & =\frac{\tau-1}{\tau}\log\frac{1}{2}-\frac{1}{\tau}\log\left(2x_{n}\right)+\frac{1}{\tau}\int_{\lambda^{-}}^{\lambda^{+}}\log\left(x_{n}+\lambda\right)\frac{\left(\lambda-\lambda^{-}\right)^{1/2}\left(\lambda^{+}-\lambda\right)^{\frac{1}{2}}}{2\pi\lambda}d\lambda\\
 & \overset{n\rightarrow\infty}{\longrightarrow}\infty.\end{align*}

Similarly, let $x_{n}\uparrow\lambda^{+}$ and $\alpha_{n}=\frac{1}{2\left(\lambda^{+}-x\right)}\in\left(0,\frac{1}{\lambda^{+}-x}\right)$.
It can be proved that as $n\rightarrow\infty$ , $\psi_{x_{n}}\left(\alpha_{n}\right)\rightarrow\infty$
and therefore $\psi_{x_{n}}^{*}\left(0\right)\rightarrow\infty$.

\item This follows from Prop. \ref{pro:Psi^Star-M} (5) and (6). 
\end{enumerate}
\end{proof}

\subsection{\label{sec:Large-Deviation-Principles}Large Deviation Principles}

This section is devoted to prove the following large deviation principle
for $\mu_{n,\mathbf{Y}}$.

\begin{thm}
\label{thm:LD-Y}Let $n,m\rightarrow\infty$ with $\frac{m}{n}\rightarrow\frac{1}{\beta}\in\mathbb{R}^{+}$.
For any $t<\bar{\lambda}-x$, \[
\underset{\left(n,m\right)\rightarrow\infty}{\lim}\frac{1}{n}\log\mu_{n,\mathbf{Y}}\left(\left.\frac{1}{n}\sum_{i=1}^{n}\left(\lambda_{i}-x\right)Y_{i}\le t\right|\bm{\lambda}\right)=-\psi_{x}^{*}\left(t\right)=-\underset{\alpha<0}{\sup}\left[\alpha t-\psi_{x}\left(\alpha\right)\right]\]
 almost surely (in $\bm{\lambda}$). Similarly, for any $t>\bar{\lambda}-x$,
\[
\underset{\left(n,m\right)\rightarrow\infty}{\lim}\frac{1}{n}\log\mu_{n,\mathbf{Y}}\left(\left.\frac{1}{n}\sum_{i=1}^{n}\left(\lambda_{i}-x\right)Y_{i}\ge t\right|\bm{\lambda}\right)=-\psi_{x}^{*}\left(t\right)=-\underset{\alpha>0}{\sup}\left[\alpha t-\psi_{x}\left(\alpha\right)\right]\]
 almost surely (in $\bm{\lambda}$). 
\end{thm}

The proof of Theorem \ref{thm:LD-Y} rests on another large deviation
principle presented in below. Recall the truncated variable $Y_{M}$
defined in (\ref{eq:u_Y_M}), its moment generating function $\psi_{x,Y_{M}}\left(\alpha\right)$
in (\ref{eq:mgf_Y_M}) and its rate function $\psi_{x,Y_{M}}^{*}\left(t\right)$
in (\ref{eq:rate_fn_Y_M}).

\begin{thm}
\label{thm:LD-Y_M}Let $n,m\rightarrow\infty$ with $\frac{m}{n}\rightarrow\frac{1}{\beta}\in\mathbb{R}^{+}$.
For any $t<\bar{\lambda}-x$ and large enough $M$,\[
\underset{\left(n,m\right)\rightarrow\infty}{\lim}\frac{1}{n}\log\mu_{n,\mathbf{Y}_{M}}\left(\left.\frac{1}{n}\sum_{i=1}^{n}\left(\lambda_{i}-x\right)Y_{M,i}\le t\right|\bm{\lambda}\right)=-\psi_{x,Y_{M}}^{*}\left(t\right)=-\underset{\alpha<0}{\sup}\left[\alpha t-\psi_{x,Y_{M}}\left(\alpha\right)\right]\]
 almost surely (in $\bm{\lambda}$). Similarly, for any $t>\bar{\lambda}-x$
and sufficiently large $M$, \[
\underset{\left(n,m\right)\rightarrow\infty}{\lim}\frac{1}{n}\log\mu_{n,\mathbf{Y}_{M}}\left(\left.\frac{1}{n}\sum_{i=1}^{n}\left(\lambda_{i}-x\right)Y_{M,i}\ge t\right|\bm{\lambda}\right)=-\psi_{x,Y_{M}}^{*}\left(t\right)=-\underset{\alpha>0}{\sup}\left[\alpha t-\psi_{x,Y_{M}}\left(\alpha\right)\right]\]
 almost surely (in $\bm{\lambda}$). 
\end{thm}

The proof of Theorem \ref{thm:LD-Y_M} is given in Appendix \ref{sub:Proof-LD-Y_M}.
Based on Theorem \ref{thm:LD-Y_M}, Theorem \ref{thm:LD-Y} is proved
in Appendix \ref{sub:Proof-LD-Y}.

\subsubsection{\label{sub:Proof-LD-Y}Proof of Theorem \ref{thm:LD-Y}}

In the following, we only prove Theorem \ref{thm:LD-Y_M} for $t<\bar{\lambda}-x$.
The $t>\bar{\lambda}-x$ case is just the dual case.

An upper bound is constructed by Chebyshev inequality. Take any $\alpha\in\left(-\frac{1}{x-\lambda_{t}^{-}},0\right)$,
\begin{align*}
 & \underset{\left(n,m\right)\rightarrow\infty}{\lim}\frac{1}{n}\log\mu_{n,\mathbf{Y}}\left(\left.\frac{1}{n}\sum_{i=1}^{n}\left(\lambda_{i}-x\right)Y_{i}\le t\right|\bm{\lambda}\right)\\
 & \le\underset{\left(n,m\right)\rightarrow\infty}{\lim}\frac{1}{n}\log\left\{ e^{-n\alpha t}\mathrm{E}_{\mathbf{Y}}\left[\left.e^{\alpha\sum_{i=1}^{n}\left(\lambda_{i}-x\right)Y_{i}}\right|\bm{\lambda}\right]\right\} \\
 & =-\left\{ \alpha t-\int\log\mathrm{E}_{Y}\left[e^{\alpha\left(\lambda-x\right)Y}\right]d\mu_{\lambda}\right\} \end{align*}
 almost surely (in $\bm{\lambda}$), where the last equality follows
from the fact that $\mathrm{E}_{Y}\left[e^{\alpha\left(\lambda-x\right)Y}\right]$
is Lipschitz on $\left[\lambda_{t}^{-},\lambda^{+}\right]$ for $\forall\alpha\in\left(-\frac{1}{x-\lambda_{t}^{-}},0\right)$.
Therefore, \begin{align}
 & \underset{\left(n,m\right)\rightarrow\infty}{\lim}\frac{1}{n}\log\mu_{n,\mathbf{Y}}\left(\left.\frac{1}{n}\sum_{i=1}^{n}\left(\lambda_{i}-x\right)Y_{i}\le t\right|\bm{\lambda}\right)\nonumber \\
 & \le-\underset{\alpha\in\left(-\frac{1}{x-\lambda_{t}^{-}},0\right)}{\sup}\left\{ \alpha t-\psi_{x}\left(\alpha\right)\right\} =-\psi_{x}^{*}\left(t\right)\label{eq:LD_Y_ub}\end{align}
 almost surely (in $\bm{\lambda}$), where the last equality follows
from Proposition \ref{pro:Psi^Star-x}(2).

A lower bound is obtained from Theorem \ref{thm:LD-Y_M}. \begin{align*}
 & \underset{\left(n,m\right)\rightarrow\infty}{\lim}\frac{1}{n}\log\mu_{n,\mathbf{Y}}\left(\left.\frac{1}{n}\sum_{i=1}^{n}\left(\lambda_{i}-x\right)Y_{i}\le t\right|\bm{\lambda}\right)\\
 & \ge\underset{\left(n,m\right)\rightarrow\infty}{\lim}\frac{1}{n}\log\mu_{n,\mathbf{Y}}\left(\left.\frac{1}{n}\sum_{i=1}^{n}\left(\lambda_{i}-x\right)Y_{i}\le t,\;\cap_{i=1}^{n}\left\{ \left|Y_{i}\right|\le M\right\} \right|\bm{\lambda}\right)\\
 & =\underset{\left(n,m\right)\rightarrow\infty}{\lim}\frac{1}{n}\log\mu_{n,\mathbf{Y}_{M}}\left(\left.\frac{1}{n}\sum_{i=1}^{n}\left(\lambda_{i}-x\right)Y_{i}\le t\right|\bm{\lambda}\right)-\log\mu_{Y}\left[0,M\right]\\
 & =-\psi_{x,Y_{M}}^{*}\left(t\right)-\log\mu_{Y}\left[0,M\right],\end{align*}
 almost surely in $\bm{\lambda}$. where the last equality follows
from Theorem \ref{thm:LD-Y_M}. Note that $\psi_{x}\left(\alpha\right)=\underset{M\rightarrow\infty}{\lim}\psi_{x,Y_{M}}\left(\alpha\right)$,
$\psi_{x}^{*}\left(t\right)=\underset{M\rightarrow\infty}{\lim}\psi_{x,Y_{M}}^{*}\left(t\right)$
and $\underset{M\rightarrow\infty}{\lim}\log\mu_{Y}\left[0,M\right]=0$.
\[
\underset{\left(n,m\right)\rightarrow\infty}{\lim}\frac{1}{n}\log\mu_{n,\mathbf{Y}}\left(\left.\frac{1}{n}\sum_{i=1}^{n}\left(\lambda_{i}-x\right)Y_{i}\le t\right|\bm{\lambda}\right)\ge-\psi_{x}^{*}\left(t\right)\]
 almost surely in $\bm{\lambda}$, which proves Theorem \ref{thm:LD-Y}.

\subsubsection{\label{sub:Proof-LD-Y_M}Proof of Theorem \ref{thm:LD-Y_M}}

The proof of this theorem follows the same line in that of Gartner-Ellis
Theorem \cite{Dembo_book92_Large_Deviation}. In the following, we
only gives the details for $t<\bar{\lambda}-x$, as the $t>\bar{\lambda}-x$
case is just the dual case.

Similar to the upper bound in (\ref{eq:LD_Y_ub}), by Chebyshev's
inequality and maximization over $\alpha\in\mathbb{R}^{-}$, we have
the upper bound \begin{align*}
 & \underset{\left(n,m\right)\rightarrow\infty}{\lim}\frac{1}{n}\log\mu_{n,\mathbf{Y}_{M}}\left(\left.\frac{1}{n}\sum_{i=1}^{n}\left(\lambda_{i}-x\right)Y_{M,i}\le t\right|\bm{\lambda}\right)\\
 & \le-\underset{\alpha<0}{\sup}\left[\alpha t-\psi_{x,Y_{M}}\left(\alpha\right)\right]\\
 & =\psi_{x,Y_{M}}^{*}\left(t\right)\end{align*}
 almost surely in $\bm{\lambda}$, where the last equality follows
from Proposition \ref{pro:Psi^Star-M}(2).

Now we prove the lower bound \begin{equation}
\underset{\left(n,m\right)\rightarrow\infty}{\lim}\frac{1}{n}\log\mu_{n,\mathbf{Y}_{M}}\left(\left.\frac{1}{n}\sum_{i=1}^{n}\left(\lambda_{i}-x\right)Y_{M,i}\le t\right|\bm{\lambda}\right)\ge-\psi_{x,Y_{M}}^{*}\left(t\right)\label{eq:LD_Y_M_lb}\end{equation}
 almost surely in $\bm{\lambda}$. This lower bound rests on the fact
(will be proved later) that for any $s\in\mathbb{R}$ and $\epsilon>0$,
\begin{equation}
\underset{\left(n,m\right)\rightarrow\infty}{\lim}\frac{1}{n}\log\mu_{n,\mathbf{Y}_{M}}\left(\left.\frac{1}{n}\sum_{i=1}^{n}\left(\lambda_{i}-x\right)Y_{M,i}\in\left(s-\epsilon,s+\epsilon\right)\right|\bm{\lambda}\right)\ge-\psi_{x,Y_{M}}^{*}\left(s\right)\label{eq:LD_Y_M_keystep}\end{equation}
 almost surely in $\bm{\lambda}$. Note that for $\forall s<t$, there
exists an $\epsilon>0$ such that $\left(s-\epsilon,s+\epsilon\right)\subset\left(-\infty,t\right)$
and \begin{align*}
 & \underset{\left(n,m\right)\rightarrow\infty}{\lim}\frac{1}{n}\log\mu_{n,\mathbf{Y}_{M}}\left(\left.\frac{1}{n}\sum_{i=1}^{n}\left(\lambda_{i}-x\right)Y_{M,i}\le t\right|\bm{\lambda}\right)\\
 & \ge\underset{\left(n,m\right)\rightarrow\infty}{\lim}\frac{1}{n}\log\mu_{n,\mathbf{Y}_{M}}\left(\left.\frac{1}{n}\sum_{i=1}^{n}\left(\lambda_{i}-x\right)Y_{M,i}\in\left(s-\epsilon,s+\epsilon\right)\right|\bm{\lambda}\right)\\
 & \ge-\psi_{x,Y_{M}}^{*}\left(s\right)\end{align*}
 almost surely in $\bm{\lambda}$. Take $s\uparrow t$. $\psi_{x,Y_{M}}^{*}\left(s\right)\rightarrow\psi_{x,Y_{M}}^{*}\left(t\right)$.
The lower bound (\ref{eq:LD_Y_M_lb}) is then proved.

The inequality (\ref{eq:LD_Y_M_keystep}) is proved by exponential
change of measure. Let $\gamma\in\mathbb{R}$ such that $\psi_{x,Y_{M}}^{*}\left(s\right)=\gamma s-\psi_{x,Y_{M}}\left(\gamma\right)$,
or equivalently, $\psi_{x,Y_{M}}^{\prime}\left(\gamma\right)=s$.
Such $\gamma$ exists according to Proposition \ref{pro:Psi^Star-M}(2).
Note that $d\mu_{n,\mathbf{y}_{M}}=\prod_{i=1}^{n}d\mu_{y_{M,i}}.$
Define a new probability measure (exponential change of $d\mu_{n,\mathbf{y}_{m}}$)
\begin{align*}
d\tilde{\mu}_{n,\mathbf{y}_{M}} & \triangleq\frac{e^{\sum_{i=1}^{n}\gamma\left(\lambda_{i}-x\right)y_{i}}}{\mathrm{E}_{\mathbf{Y}_{M}}\left[e^{\sum_{i=1}^{n}\gamma\left(\lambda_{i}-x\right)Y_{M,i}}\right]}d\mu_{n,\mathbf{y}_{M}}\\
 & =\frac{e^{\sum_{i=1}^{n}\gamma\left(\lambda_{i}-x\right)y_{i}}}{e^{\sum_{i=1}^{n}\log\mathrm{E}_{Y_{M}}\left[e^{\gamma\left(\lambda_{i}-x\right)Y_{M}}\right]}}d\mu_{n,\mathbf{y}_{M}}.\end{align*}
 Since $\int d\tilde{\mu}_{n,\mathbf{y}_{M}}=1$, $\tilde{\mu}_{n,\mathbf{y}_{M}}$
is a well defined probability measure. Let \[
A_{n,\mathbf{y}}\triangleq\left\{ \left(y_{1},\cdots,y_{n}\right):\;\mathbf{y}\in\prod_{i=1}^{n}\left[0,M\right],\;\frac{1}{n}\sum_{i=1}^{n}\left(\lambda_{i}-x\right)y_{i}\in\left(s-\epsilon,s+\epsilon\right)\right\} .\]
 Then\begin{align*}
 & \frac{1}{n}\log\mu_{n,\mathbf{Y}_{M}}\left(\left.\sum_{i=1}^{n}\left(\lambda_{i}-x\right)Y_{M,i}\in\left(s-\epsilon,s+\epsilon\right)\right|\bm{\lambda}\right)\\
 & =\frac{1}{n}\log\int_{A_{n,\mathbf{y}}}d\mu_{n,\mathbf{y}_{M}}\\
 & =\frac{1}{n}\log\int_{A_{n,\mathbf{y}}}\left(e^{\sum_{i=1}^{n}\log\mathrm{E}_{Y_{M}}\left[\left.e^{\gamma\left(\lambda_{i}-x\right)Y_{M}}\right|\bm{\lambda}\right]}\cdot e^{-\sum_{i=1}^{n}\gamma\left(\lambda_{i}-x\right)y_{i}}\right)d\tilde{\mu}_{n,\mathbf{y}_{M}}\\
 & =\frac{1}{n}\log e^{-n\left\{ \gamma s-\frac{1}{n}\sum_{i=1}^{n}\log\mathrm{E}_{Y_{M}}\left[\left.e^{\gamma\left(\lambda_{i}-x\right)Y_{M}}\right|\bm{\lambda}\right]\right\} }\\
 & \quad\quad+\frac{1}{n}\log\int_{A_{n,\mathbf{y}}}e^{-n\gamma\left[\frac{1}{n}\sum_{i=1}^{n}\left(\lambda_{i}-x\right)y_{i}-s\right]}d\tilde{\mu}_{n,\mathbf{y}_{M}}\\
 & \ge-\left\{ \gamma s-\frac{1}{n}\sum_{i=1}^{n}\log\mathrm{E}_{Y_{M}}\left[\left.e^{\gamma\left(\lambda_{i}-x\right)Y_{M}}\right|\bm{\lambda}\right]\right\} +\frac{1}{n}\log\left(e^{-n\left|\gamma\epsilon\right|}\int_{A_{n,\mathbf{y}}}d\tilde{\mu}_{n,\mathbf{y}_{M}}\right)\\
 & =-\left\{ \gamma s-\frac{1}{n}\sum_{i=1}^{n}\log\mathrm{E}_{Y_{M}}\left[\left.e^{\gamma\left(\lambda_{i}-x\right)Y_{M}}\right|\bm{\lambda}\right]\right\} -\left|\gamma\epsilon\right|+\frac{1}{n}\log\tilde{\mu}_{n,\mathbf{Y}_{M}}\left(\left.A_{n,\mathbf{y}}\right|\bm{\lambda}\right).\end{align*}
 Note that $\underset{\left(n,m\right)\rightarrow\infty}{\lim}\;\frac{1}{n}\sum_{i=1}^{n}\log\mathrm{E}_{Y_{M}}\left[\left.e^{\gamma\left(\lambda_{i}-x\right)Y_{M}}\right|\bm{\lambda}\right]=\psi_{x,Y_{M}}\left(\gamma\right)$
almost surely in $\bm{\lambda}$. (\ref{eq:LD_Y_M_keystep}) is true
if \begin{equation}
\underset{\left(n,m\right)\rightarrow\infty}{\lim}\frac{1}{n}\log\tilde{\mu}_{n,\mathbf{Y}_{M}}\left(\left.A_{n,\mathbf{y}}\right|\bm{\lambda}\right)=0\label{eq:shifted_measure_A_n}\end{equation}
 almost surely in $\bm{\lambda}$.

In order to prove (\ref{eq:shifted_measure_A_n}), note that \begin{align*}
\tilde{\mu}_{n,\mathbf{Y}_{M}}\left(\left.A_{n,\mathbf{y}}\right|\bm{\lambda}\right) & =1-\tilde{\mu}_{n,\mathbf{Y}_{M}}\left(\left.\frac{1}{n}\sum_{i=1}^{n}\left(\lambda_{i}-x\right)y_{i}\le s-\epsilon\right|\bm{\lambda}\right)\\
 & \quad\quad-\tilde{\mu}_{n,\mathbf{Y}_{M}}\left(\left.\frac{1}{n}\sum_{i=1}^{n}\left(\lambda_{i}-x\right)y_{i}\ge s+\epsilon\right|\bm{\lambda}\right).\end{align*}
 We upper bound $\tilde{\mu}_{n,\mathbf{Y}_{M}}\left(\left.\frac{1}{n}\sum_{i=1}^{n}\left(\lambda_{i}-x\right)y_{i}\le s-\epsilon\right|\bm{\lambda}\right)$
and $\tilde{\mu}_{n,\mathbf{Y}_{M}}\left(\left.\frac{1}{n}\sum_{i=1}^{n}\left(\lambda_{i}-x\right)y_{i}\ge s+\epsilon\right|\bm{\lambda}\right)$
respectively. Note that \begin{align*}
 & \underset{\left(n,m\right)\rightarrow\infty}{\lim}\frac{1}{n}\log\int e^{\alpha\sum\left(\lambda_{i}-x\right)y_{i}}d\tilde{\mu}_{n,\mathbf{y}_{M}}\\
 & =\underset{\left(n,m\right)\rightarrow\infty}{\lim}\left[\frac{1}{n}\log\int e^{\left(\alpha+\gamma\right)\sum\left(\lambda_{i}-x\right)y_{i}}d\mu_{n,\mathbf{y}_{M}}-\frac{1}{n}\log\mathrm{E}_{\mathbf{Y}_{M}}\left[\left.e^{\gamma\sum\left(\lambda_{i}-x\right)Y_{M,i}}\right|\bm{\lambda}\right]\right]\\
 & =\psi_{x,Y_{M}}\left(\alpha+\gamma\right)-\psi_{x,Y_{M}}\left(\gamma\right)\end{align*}
 almost surely in $\bm{\lambda}$. Define $\tilde{\psi}_{x,Y_{M}}\left(\alpha\right)=\psi_{x,Y_{M}}\left(\alpha+\gamma\right)-\psi_{x,Y_{M}}\left(\gamma\right)$
and $\tilde{\psi}_{x,Y_{M}}^{*}\left(s\right)=\underset{\alpha}{\sup}\left[\alpha s-\tilde{\psi}_{x,Y_{M}}\left(\alpha\right)\right]$.
Since $\psi_{x,Y_{M}}\left(\alpha\right)$ is strictly convex (Proposition
\ref{pro:Psi-Properties}(4)), $\tilde{\psi}_{x,Y_{M}}\left(\alpha\right)$
is strictly convex. Note that $\psi_{x,Y_{M}}^{\prime}\left(\gamma\right)=s$
as $M$ is sufficiently large. For large enough $M$, \[
\tilde{\psi}_{x,Y_{M}}^{\prime}\left(\alpha\right)=\psi_{x,Y_{M}}^{\prime}\left(\alpha+\gamma\right)\begin{cases}
>s & \mathrm{if}\;\alpha>0\\
=s & \mathrm{if}\;\alpha=0\\
<s & \mathrm{if}\;\alpha<0\end{cases}.\]
 Therefore, $\tilde{\psi}_{x,Y_{M}}^{*}\left(s-\epsilon\right)$ is
achieved at an $\alpha_{s-\epsilon}<0$ and \begin{align*}
\tilde{\psi}_{x,Y_{M}}^{*}\left(s-\epsilon\right) & =\alpha_{s-\epsilon}\left(s-\epsilon\right)-\tilde{\psi}_{x,Y_{M}}\left(\alpha_{s-\epsilon}\right)\\
 & =\left(\alpha_{s-\epsilon}+\gamma\right)\left(s-\epsilon\right)-\psi_{x,Y_{M}}\left(\alpha_{s-\epsilon}+\gamma\right)-\gamma s+\psi_{x,Y_{M}}\left(\gamma\right)+\gamma\epsilon\\
 & \overset{\left(a\right)}{=}\gamma\epsilon-\left[\psi_{x,Y_{M}}^{*}\left(s\right)-\psi_{x,Y_{M}}^{*}\left(s-\epsilon\right)\right]\\
 & \overset{\left(b\right)}{>}0,\end{align*}
 where $\left(a\right)$ is from the fact that $\psi_{x,Y_{M}}^{\prime}\left(\alpha_{t-\epsilon}+\gamma\right)=t-\epsilon$,
and $\left(b\right)$ follows from Proposition \ref{pro:Psi^Star-M}(3).
Similarly, $\tilde{\psi}_{x,Y_{M}}^{*}\left(s+\epsilon\right)$ is
achieved at an $\alpha_{s+\epsilon}>0$ and $\tilde{\psi}_{x}^{*}\left(ts+\epsilon\right)>0$.
Now, by Chebyshev's inequality, \[
\underset{\left(n,m\right)\rightarrow\infty}{\lim}\frac{1}{n}\log\tilde{\mu}_{n,\mathbf{Y}_{M}}\left(\left.\frac{1}{n}\sum\left(\lambda_{i}-x\right)\tilde{Y}_{M,i}\le s-\epsilon\right|\bm{\lambda}\right)=-\tilde{\psi}_{x,Y_{M}}^{*}\left(s-\epsilon\right)<0,\]
 and\[
\underset{\left(n,m\right)\rightarrow\infty}{\lim}\frac{1}{n}\log\tilde{\mu}_{n,\mathbf{Y}_{M}}\left(\left|\frac{1}{n}\sum\left(\lambda_{i}-x\right)\tilde{Y}_{M,i}\le s-\epsilon\right|\bm{\lambda}\right)=-\tilde{\psi}_{x,Y_{M}}^{*}\left(s+\epsilon\right)<0.\]
 almost surely in $\bm{\lambda}$. Take a $b>0$ such that $\tilde{\psi}_{x}^{*}\left(t-\epsilon\right)>b$
and $\tilde{\psi}_{x}^{*}\left(t+\epsilon\right)>b$. Then, \begin{align*}
 & \underset{\left(n,m\right)\rightarrow\infty}{\lim}\frac{1}{n}\log\tilde{\mu}_{n,\mathbf{Y}_{M}}\left(\left.A_{n,\mathbf{y}}\right|\bm{\lambda}\right)\\
 & \ge\underset{\left(n,m\right)\rightarrow\infty}{\lim}\frac{1}{n}\log\left(1-2e^{-nb\left(1+o\left(1\right)\right)}\right)=0,\end{align*}
 which is exactly (\ref{eq:shifted_measure_A_n}).

\subsection{\label{sec:Computation}Proof of Corollary \ref{cor:main-result-computations}}

The following lemma, proved in \cite{Verdu_IT1999_CDMA_Spectral_Efficiency},
is essential to our derivation in Corollary \ref{cor:main-result-computations}.

\begin{lemma}
\label{lem:Verdu-Result}For $\forall z\in\mathbb{C}\backslash\left[\lambda^{-},\lambda^{+}\right]$,
\begin{equation}
\int_{0}^{\infty}\frac{z\lambda}{1+z\lambda}f_{\beta}\left(\lambda\right)d\lambda=\frac{\mathcal{F}\left(z,\beta\right)}{4z\beta},\label{eq:derivation-formula}\end{equation}
 and \begin{equation}
\int_{0}^{\infty}\log\left(1+z\lambda\right)f_{\beta}\left(\lambda\right)d\lambda=\log\left(1+z-\frac{1}{4}\mathcal{F}\left(z,\beta\right)\right)+\frac{1}{\beta}\log\left(1+z\beta-\frac{1}{4}\mathcal{F}\left(z,\beta\right)\right)-\frac{\mathcal{F}\left(z,\beta\right)}{4z\beta},\label{eq:throughput-formula}\end{equation}
 where \[
\mathcal{F}\left(z,\beta\right)\triangleq\left(\left(1+\lambda^{-}z\right)^{1/2}-\left(1+\lambda^{+}z\right)^{1/2}\right)^{2}.\]
 
\end{lemma}

The basic step uses (\ref{eq:derivation-formula}) to identify the
$\alpha^{*}$ at which $\psi_{x}^{*}\left(0\right)$ is achieved.

\begin{prop}
\label{pro:optimal_alpha}Let $x\in\left(\lambda_{t}^{-},\lambda^{+}\right)$.
Let $\alpha^{*}$ be such that $\psi_{x}^{*}\left(0\right)=\int\log\left(1+\alpha^{*}\left(x-\lambda\right)\right)d\mu_{\lambda}$.
Then \[
\alpha^{*}=\begin{cases}
\frac{1}{\lambda^{+}-x} & \mathrm{if}\; x\ge1+\sqrt{\beta},\\
-\frac{1}{x-\lambda^{-}} & \mathrm{if}\; x\le1-\sqrt{\beta}\;\mathrm{and}\;\beta<1,\\
\frac{1}{\beta}\frac{x-1}{x} & \mathrm{otherwise}.\end{cases}\]
 
\end{prop}
\begin{proof}
Since $-\psi_{x}\left(\alpha\right)=\int\log\left(1+\alpha\left(x-\lambda\right)\right)d\mu_{\lambda}$
is concave on $\alpha\in\left(-\frac{1}{x-\lambda_{t}^{-}},\frac{1}{\lambda^{+}-x}\right)$
(Proposition \ref{pro:Psi-Properties}(4)), $\alpha^{*}$ can be found
by evaluating $-\psi_{x}^{\prime}\left(\alpha\right)$.

We computes $-\psi_{x}^{\prime}\left(\alpha\right)$ for any $\alpha\neq-\frac{1}{x},$
\begin{align}
 & \int\frac{x-\lambda}{1+\alpha\left(x-\lambda\right)}d\mu_{\lambda}\nonumber \\
 & =\frac{1}{\alpha}\left[\frac{\alpha x}{1+\alpha x}+\frac{1}{1+\alpha x}\int\frac{-\frac{\alpha}{1+\alpha x}\lambda}{1-\frac{\alpha}{1+\alpha x}\lambda}d\mu_{\lambda}\right]\nonumber \\
 & =\frac{1}{\alpha}\left[\frac{\alpha x}{1+\alpha x}-\frac{\mathcal{F}\left(-\frac{\alpha}{1+\alpha x},\beta\right)}{4\alpha\beta}\right],\label{eq:derivative-psi_x}\end{align}
 where the last line follows from (\ref{eq:derivation-formula}).

By evaluating $\psi_{x}^{\prime}\left(\alpha\right)$ at the boundary
points $\alpha=\frac{1}{\lambda^{+}-x}$ and $\alpha=-\frac{1}{x-\lambda_{t}^{-}}$,
it can be verified that $\alpha^{*}$ satisfies\[
\begin{cases}
\alpha^{*}=-\frac{1}{x-\lambda^{-}} & \mathrm{if}\; x<1-\sqrt{\beta}\;\mathrm{and}\;\beta<1\\
\alpha^{*}=\frac{1}{\lambda^{+}-x} & \mathrm{if}\; x>1+\sqrt{\beta}\\
\psi_{x}^{\prime}\left(\alpha^{*}\right)=0 & \mathrm{otherwise}\end{cases}.\]

We shall find $\alpha\in\left(-\frac{1}{x-\lambda_{t}^{-}},\frac{1}{\lambda^{+}-x}\right)$
such that $\psi_{x}^{\prime}\left(\alpha\right)=0$. Suppose that
$\alpha\neq-\frac{1}{x}$ where $\psi_{x}^{\prime}\left(\alpha\right)$
can be computed from (\ref{eq:derivative-psi_x}). $\psi_{x}^{\prime}\left(\alpha\right)=0$
implies $\mathcal{F}\left(-\frac{\alpha}{1+\alpha x},\beta\right)=\frac{4\alpha^{2}\beta x}{1+\alpha x}.$
Let $y=\alpha\beta$. Then \begin{align*}
\frac{4\alpha^{2}\beta x}{1+y} & =\frac{2}{1+y}\left(1+y-\alpha\left(1+\beta\right)-2\sqrt{\left(\left(1+y\right)-\alpha\left(1+\beta\right)\right)^{2}-4\alpha^{2}\beta}\right).\end{align*}
 Elementary simplification gives a quadratic equation\[
\left(\alpha\beta-1\right)y^{2}+\left(\alpha+\alpha\beta-1\right)y+\alpha=0,\]
 whose two roots $y_{1}=-1$ and $y_{2}=\frac{\alpha}{1-\alpha\beta}$.
Since we have assumed that $\alpha\neq-\frac{1}{x}$, the only possible
root is $y_{2}=\frac{\alpha}{1-\alpha\beta}$. It is then clear that
\begin{equation}
\alpha^{*}=\frac{1}{\beta}\frac{x-1}{x}.\label{eq:optimal-alpha}\end{equation}

Finally, we shall discuss the case that $\alpha=-\frac{1}{x}$. $\psi_{x}^{\prime}\left(-\frac{1}{x}\right)=0$
implies that $\beta<1$ and $x\in\left(1-\sqrt{\beta},1\right)$.
Note that \begin{align*}
0 & =\int\frac{x-\lambda}{1+\left(-\frac{1}{x}\right)\left(x-\lambda\right)}d\mu_{\lambda}\\
 & =x\left[x\int\frac{1}{\lambda}d\mu_{\lambda}-1\right]\\
 & =x\left[x\frac{1}{1-\beta}-1\right].\end{align*}
 We obtain that $x=1-\beta$. However, $\left.\frac{1}{\beta}\frac{x-1}{x}\right|_{x=1-\beta}=-\frac{1}{1-\beta}=-\frac{1}{x}$.
The case $\alpha=-\frac{1}{x}$ is a special case of (\ref{eq:optimal-alpha}). 
\end{proof}

Now for any given $r\in\mathbb{R}^{+}$, we compute $x_{r}^{-}$ and
$x_{r}^{+}$. Note that $\psi_{x}^{*}\left(0\right)=-\psi_{x}\left(\alpha^{*}\right)$.
By Proposition \ref{pro:optimal_alpha} and (\ref{eq:throughput-formula}),
we are able to solve the equation $r\log2=-\psi_{x}\left(\alpha^{*}\right)$.
The results are presented in Corollary \ref{cor:main-result-computations}.

 \bibliographystyle{IEEEtran}
\bibliography{bib/_Dai,bib/Honig,bib/Math}
 
\end{document}